\documentclass[final,authoryear,3p,times,twocolumn]{elsarticle}

\usepackage[latin2]{inputenc}
\usepackage{graphicx}

\usepackage{amssymb}
\usepackage{amsthm}
\usepackage{ulem}

\journal{Applied Radiation and Isotopes}

\begin{document}

\begin{frontmatter}



\title{New activation cross section data on longer lived radio-nuclei produced in proton induced nuclear reaction on zirconium}


\author[1]{F. T\'ark\'anyi}
\author[1]{F. Ditr\'oi\corref{*}}
\author[1]{S. Tak\'acs}
\author[2]{A. Hermanne}
\author[3]{M. Al-Abyad}
\author[4]{H. Yamazaki}
\author[4]{M. Baba}
\author[4]{M.A. Mohammadi}

\cortext[*]{Corresponding author: ditroi@atomki.hu}

\address[1]{Institute for Nuclear Research, Hungarian Academy of Sciences (ATOMKI),  Debrecen, Hungary}
\address[2]{Cyclotron Laboratory, Vrije Universiteit Brussel (VUB), Brussels, Belgium}
\address[3]{Physics Department, Cyclotron Facility, Nuclear Research Centre, Atomic Energy Authority, Cairo 13759, Egypt}
\address[4]{Cyclotron and Radioisotope Center (CYRIC), Tohoku University, Sendai, Japan}

\begin{abstract}
In the frame of a systematic study of charged particle production routes of medically relevant radionuclei, the excitation function for indirect production of $^{178m}$Ta through $^{nat}$Hf($\alpha$,xn)$^{178-178m}$Ta nuclear reaction was measured for the first time up to 40 MeV. In parallel, the side reactions $^{nat}$Hf($\alpha$,x)$^{179,177,176,175}$W, $^{183,182,178g,177,176,175}$Ta, $^{179m,177m,175}$Hf were also assessed. Stacked foil irradiation technique and $\gamma$-ray spectrometry were used. New experimental cross section data for the $^{nat}$Ta(d,xn)$^{178}$W reaction are also reported up to 40 MeV.  The measured excitation functions are compared with the results of the ALICE-IPPE, and EMPIRE nuclear reaction model codes and with the TALYS 1.4 based data in the TENDL-2013 library. The thick target yields were deduced and compared with yields of other charged particle ((p,4n), (d,5n) and ($^3$He,x)) production routes for $^{178}$W.
\end{abstract}


\end{frontmatter}


\section{Introduction}
\label{1}
Production cross sections of proton induced nuclear reactions on metals are important for many applications and for development of improved nuclear reaction theory. In most applications high intensity, low and high energy direct or secondary proton beams activate technological elements and produce highly active radio-products. After recognizing the importance of knowledge of production cross sections we concluded that a systematic coordinated experimental and theoretical study is necessary, and we started a set of experiments with a large scope.  
Our research in connection with the activation cross sections on zirconium is of importance and applies to different projects: 
\begin{itemize}
\item	 Preparation of a nuclear database for production of $^{90}$Nb, $^{95}$Nb, $^{89}$Zr, $^{88}$Y medical radioisotopes in the frame of IAEA Coordinated Research Project \citep{Gul, IAEA2001, IAEA2012} using the $^{90}$Zr(p,n)$^{90}$Nb, $^{96}$Zr(p,2n)$^{95}$Nb, $^{90}$Zr(p,2n)$^{89}$Nb-$^{89}$Zr and $^{nat}$Zr(p,x)$^{88}$Zr$\longrightarrow$ $^{88}$Y production routes.
We have also investigated alternative production routes of these radio-products on yttrium \citep{Uddin2007, Uddin2005}.
\item	Preparation of proton and deuteron activation cross section database for the Fusion Evaluated Nuclear Data Library \citep{IAEA2004}.
\item	Preparation of a database for the Thin Layer Activation (TLA) technique for wear measurement \citep{IAEA2010} and every day practice of wear measurement of zirconium alloy samples.
\end{itemize}
We earlier reported on experimental activation cross section data on Zr targets for deuteron induced reactions up to 50 MeV \citep{TF2004, TF1}  and for proton reactions up to 17 MeV \citep{Abyad}. During the compilation of the experimental data of proton induced activation a large disagreements in the database at higher energies were noted. Hence we decided to extend our investigations up to higher energies and to complete them by testing the prediction capabilities of the widely used TALYS model code.

\section{REVIEW OF EARLIER INVESTIGATIONS}
\label{2}
\subsection{Earlier experimental investigations}
\label{2.1}
The earlier experimental investigations reported in the literature (also from our group) are compiled and shown in Table 1 that includes information on the used target, accelerator, beam monitoring, measurement of the activity, the measured data points, and the covered energy range. According to our tradition we have investigated the earlier experimental results and the theoretical results in detail before new experiments were designed and performed. 

\begin{table*}[t]
\tiny
\caption{Summary of earlier experimental investigations on activation cross sections and yields of the proton induced nuclear reaction on zirconium. The investigated quantities of the nuclear reactions are indicated according to the conventions of the EXFOR system \citep{IAEA2008}.}
\begin{center}
\begin{tabular}{|p{0.6in}|p{0.5in}|p{0.8in}|p{1.in}|p{0.8in}|p{1.4in}|p{0.4in}|} 
\hline
\textbf{Author} & \textbf{Target} & \textbf{Irradiation} & 
\textbf{Beam current measurement and monitor reaction} & \textbf{
Separation method and measurement of activity} & \textbf{Nuclear 
reaction and measured quantity and number of measured data points } & 
\textbf{Covered energy range(MeV)} \\
\hline
\citep{Blaser} & $^{nat}$Zr nitrate & cyclotron stacked 
foil & $^{63}$Cu(p,n)$^{63}$Zn\newline  $^{62}$Ni(p,n)$^{62}$Cu & 
Geiger-M\"uller counter & 40-ZR-96(P,N)41-NB-96,SIG, 
15\newline 40-ZR-92(P,N)41-NB-92,SIG, 10\newline 40-ZR-91(P,N)41-NB-91-M,SIG, 12 & 
2.73-6.67\newline 3.5-6.67\newline 3.5-6.66 \\
\hline
\citep{Blosser} & single foil & cyclotron & $^{63
}$Cu(p,n)$^{63}$Zn & Geiger-M\"uller counter & 
40-ZR-90(P,N)41-NB-90-G,M+,SIG,EXP,1 & 12.7 \\
\hline
\citep{Delaunau} & $^{90}$Zr$^{91
}$Zr$^{94}$Zr & cyclotron & $^{65}$Cu(p,n)$^{65}$Zn & 
$\gamma$-NaI(Tl) & 40-ZR-90(P,$\alpha$)39-Y-87,SIG,EXP, 1\newline 40-ZR-91(P,$\alpha$)39-Y-88,SIG,EXP,
1\newline 40-ZR-94(P,$\alpha$)39-Y-91,SIG,EXP, 1 & 11.2\newline 11.2\newline 11.2 \\
\hline
\citep{Kantelo} & $^{90}$Zr-oxide & cyclotron single 
target & $^{63}$Cu(p,x)$^{63}$Zn\newline $^{65}$Cu(p,x)$^{64}$Cu & 
$\gamma$-Ge (Li) & 
40-ZR-90(P,X)39-Y-86-G,M+,SIG,EXP,1\newline 40-ZR-90(P,X)39-Y-87-G,M+,SIG,EXP,1\newline 
40-ZR-90(P,X)39-Y-88, SIG,EXP,1\newline 40-ZR-90(P,X)39-Y-85-G,SIG,EXP\newline 40-ZR-90(P,X)39-Y-85-M,SIG,EXP, 
1 & 11.2\newline 11.2\newline 11.2\newline 11.2 \\
\hline
\citep{Birjukov} & $^{91}$Zr(63.63 \%) & 
cyclotron single target & beam current integrator & neutron time of flight 
& 40-ZR-90(P,N)41-NB-90,SIG, 1 & 11.2 \\
\hline
\citep{Roughton} & $^{nat}$Zr & cyclotron single target 
irr. & beam current integrator & $\gamma$-Ge(Li) & 40-ZR-90(P,G)41-NB-91-M,PY, 
TT, 15\newline 40-ZR-94(P,N)41-NB-94-M,PY,TT 13\newline 40-ZR-96(P,N)41-NB-96,PY,TT, 12 & 
1.75-6.426\newline 3.056-5.953\newline 3.057-5.954 \\
\hline
\citep{Muminov} & $^{nat}$Zr & cyclotron & & no 
chemical separation Ge(Li) & 40-ZR-0(P,N)41-NB-90-M,TTY, 3 & 9-11 \\
\hline
\citep{Dmitriev81} & $^{nat}$Zr & cyclotron single 
target & Faraday cup & $\gamma$-Ge(Li) & 40-ZR-0(P,X)39-Y-88,TTY,DT, 
1\newline 40-ZR-0(P,X)40-ZR-95,CUM,TTY,DT, 1\newline 40-ZR-0(P,X)41-NB-92-M,TTY,DT, 
1\newline 40-ZR-96(P,2N)41-NB-95-G,TTY,DT, 1 & 22\newline 22\newline 22\newline 22 \\
\hline
\citep{Regnier}) & $^{nat}$Zr & 
cyclotron linac cyclotron stacked foil & $^{27}$Al(p,x)$^{22}$Na & 
$\gamma$--HPGe & 40-ZR-0(P,X)40-ZR-88,CUM,SIG, 7 & 59-24000 \\
\hline
\citep{Dmitriev83} & & & & & 40-ZR-92(P,N)41-NB-92-M,TTY,EXP, 
140-ZR-92(P,2N)41-NB-91-M,TTY,EXP, 1\newline 40-ZR-96(P,2N)41-NB-95,TTY,EXP, 1 \newline 
40-ZR-90(P,X)40-ZR-89,TTY,EXP, 1\newline 40-ZR-90(P,A)39-Y-87,TTY,EXP, 
1\newline 40-ZR-96(P,X)40-ZR-95,TTY,EXP, 1 & 22\newline 22\newline 22\newline 22\newline 22\newline 22 \\
\hline
\citep{Abe} & $^{nat}$Zr & cyclotron multi-sample target 
method & $^{65}$Cu(p,n)$^{65}$Zn & $\gamma$-Ge(Li), HPGe & 
40-ZR-0(P,X)41-NB-90-G,M+,TTY,DT, 1\newline 40-ZR-92(P,N)41-NB-92-M,TTY,DT, 
1\newline 40-ZR-96(P,N)41-NB-96,TTY,DT, 1 & 16\newline 16\newline 16 \\
\hline
\citep{Isshiki}) & $^{nat}$Zr & cyclotron rotating target 
holder & $^{nat}$Ti(p,x)$^{48}$V & ?-Ge(Li) & 
40-ZR-0(P,N)41-NB-90-M,TTY, 1 & 10.4 \\
\hline
\citep{Batij} & $^{96}$Zr & LINAC single foil irradiation 
& Faraday cup & $\gamma$-Ge(Li) & 40-ZR-90(P,N)41-NB-90-M,SIG, 4 \newline
40-ZR-90(P,N)41-NB-90-G,SIG, 3\newline 40-ZR-90(P,N)41-NB-90,SIG, 
3\newline 40-ZR-92(P,N)41-NB-92-M,SIG, 10\newline 40-ZR-96(P,N)41-NB-96,SIG, 11 & 
7.5-9\newline 8-9\newline 8-9\newline 4.5-9\newline 4-9 \\
\hline
\citep{Konstantinov} & $^{nat}$Zr & 
syclotron stacked foil & Faraday cup & $\gamma$-Ge(Li)) & 
40-ZR-0(P,X)39-Y-87-G,,TTY,(PHY), 28\newline 40-ZR-0(P,X)39-Y-88,,TTY,(PHY), 
30\newline 40-ZR-0(P,X)40-ZR-89-G,TTY,(PHY), 16\newline 40-ZR-0(P,X)41-NB-91-M,TTY,(PHY), 
18\newline 40-ZR-0(P,X)41-NB-92-M,TTY,(PHY), 35\newline 40-ZR-0(P,X)41-NB-95-G,TTY,(PHY), 
30\newline 40-ZR-0(P,X)39-Y-87-G, TTY,(PHY), 1\newline 40-ZR-0(P,X)39-Y-88,TTY,(PHY), 
1\newline 40-ZR-0(P,X)40-ZR-89-G,TTY,(PHY), 1\newline 40-ZR-0(P,X)40-ZR-95,TTY,(PHY), 
1\newline 40-ZR-0(P,X)41-NB-91-M,TTY,(PHY), 1\newline 40-ZR-0(P,X)41-NB-92-M,TTY,(PHY), 
1\newline 40-ZR-0(P,X)41-NB-95-M,TTY,(PHY), 1\newline 40-ZR-0(P,X)41-NB-95-G,TTY,(PHY), 1 
& 
11-22.4\newline 9.8-22.4\newline 16.8-22.4\newline 15.9-22.4\newline 6.1-22.4\newline 9.8-22.4\newline 22.4 22.4\newline 22.4 22.4\newline 22.4\newline 22.4\newline 22.4\newline 22.4 
\\
\hline
\citep{Kuzmenko} & $^{nat}$Zr & cyclotron stacked foil 
& Faraday cup & $\gamma$-Ge(Li) & 40-ZR-90(P,N)41-NB-90-G,SIG, 
1\newline 40-ZR-0(P,X)40-ZR-89,CUM,SIG, 1\newline 40-ZR-0(P,X)39-Y-87,SIG, 
1\newline 40-ZR-0(P,X)39-Y-86,SIG, 1 & 17\newline 19\newline19\newline28 \\
\hline
\citep{Skakun} & $^{90}$Zr$^{92}$Zr$^{96}$Zr & 
LINACsingle foil target & Faraday cup & $\gamma$-Ge(Li) & 
40-ZR-90(P,N)41-NB-90-M,,SIG, 4\newline 40-ZR-90(P,N)41-NB-90-G,SIG, 
3\newline 40-ZR-90(P,N)41-NB-90,SIG, 3\newline 40-ZR-92(P,N)41-NB-92-M,SIG,10\newline 
40-ZR-96(P,N)41-NB-96,SIG, 11 & 7.5-9\newline 8-9\newline 8-9\newline 4.5-94-9 \\
\hline
\citep{Gorpinich} & $^{nat}$Zr & cyclotron stacked 
foil & Faraday cup & $\gamma$-Ge(Li) & 40-ZR-0(P,X)37-RB-82-M,SIG,EXP, 
18\newline 40-ZR-0(P,X)37-RB-83, SIG,EXP,24\newline 40-ZR-0(P,X)37-RB-84,SIG,EXP, 14 & 
43.9-68\newline 35-68\newline 47.5-68 \\
\hline
\citep{Nickles} & $^{nat}$Zr & cyclotron & Faraday cup & 
$\gamma$-Ge(Li) & 40-ZR-90(P,N)41-NB-90, TTY, 1\newline 40-ZR-92(P,N)41-NB-92-M,TTY, 
1\newline 40-ZR-96(P,N)41-NB-96,TTY, 1 & 111111 \\
\hline
\citep{Kondratev} & $^{nat}$Zr & cyclotron & $^{27
}$Al(p,x)$^{24}\newline $Na$^{nat}$Fe(p,x)$^{56} $Co$^{nat}$
Fe(p,x)$^{51}$Cr & $\gamma$-Ge(Li) & 40-ZR-0(P,X)41-NB-90-G,IND/M+,SIG, 39 
& 9.6-68.0 \\
\hline
\citep{Vysotsky} & $^{nat}$Zr & cyclotron stacked foil 
& & $\gamma$-Ge(Li) & 40-ZR-0(P,X)39-Y-88,(CUM),SIG,EXP, 
640-ZR-0(P,X)41-NB-91-M,IND,SIG,EXP, 5\newline 40-ZR-0(P,X)41-NB-92-M,SIG,EXP, 
5\newline 40-ZR-0(P,X)41-NB-95-G,(CUM),SIG,EXP, 6 & 1.6-5.7\newline 2.8-5.7\newline 2.8-5.7\newline1.6-5.7 
\\
\hline
\citep{Levkovskii} & $^{90}$Zr$^{91}$Zr$^{92}$Zr
$^{96}$Zr & cyclotron rotating target & $^{nat}$Mo(p,x)$^{96}$
Tc & no chemical separation $\gamma$-Ge (Li) & 40-ZR-90(P,2N)41-NB-89-M,,SIG, 
16\newline 40-ZR-90(P,$\alpha$)39-Y-87-G,SIG, 13\newline 40-ZR-90(P,$\alpha$)39-Y-87-M,SIG, 
16\newline 40-ZR-90(P,N)41-NB-90,SIG, 25\newline 40-ZR-90(P,N+$\alpha$)39-Y-86,SIG, 
9\newline 40-ZR-90(P,N+$\alpha$)39-Y-86-M,SIG, 
7\newline 40-ZR-90(P,X)40-ZR-89,CUM,SIG\newline 40-ZR-90(P,N+P)40-ZR-89,SIG+40-ZR-90(P,2N)41-NB-89-M,SIG, 
20\newline 40-ZR-91(P,2N)41-NB-90,SIG, 17\newline 40-ZR-91(P,$\alpha$)39-Y-88,SIG, 
21\newline 40-ZR-91(P,N+$\alpha$)39-Y-87-G,SIG, 8\newline 40-ZR-91(P,N+$\alpha$)39-Y-87-M,SIG, 
10\newline 40-ZR-91(P,X)40-ZR-89,CUM,SIG\newline 40-ZR-91(P,2N+P)40-ZR-89,SIG\newline 40-ZR-91(P,3N)41-NB-89-M,SIG, 
8\newline 40-ZR-92(P,3N)41-NB-90,SIG, 8\newline 40-ZR-92(P,N+$\alpha$)39-Y-88,SIG, 
13\newline 40-ZR-96(P,2N)41-NB-95-G,SIG, 25\newline 40-ZR-96(P,2N)41-NB-95-M,SIG, 
25\newline 40-ZR-96(P,N)41-NB-96,SIG, 13\newline 40-ZR-96(P,N+P)40-ZR-95,SIG, 20 & 
15.8-29.5\newline 18.3-29.5\newline 15.8-29.5\newline 7.7-29.5\newline 22.7-29.5\newline 24-29.5\newline\newline\newline 12.1-29.5\newline 14.8-29.5\newline 9.5-27.6\newline 23.1-29.5\newline 21.4-29.5\newline\newline 23.1-29.5\newline 23.1-29.5\newline 18.3-29.5\newline 7.7-29.5\newline 7.7-29.5\newline 7.7-18.3\newline 12.1-29.5 
\\
\hline

\end{tabular}

\end{center}

\end{table*}

\setcounter{table}{0}
\begin{table*}[t]
\tiny
\caption{continued}
\begin{center}
\begin{tabular}{|p{0.6in}|p{0.5in}|p{0.8in}|p{1.in}|p{0.8in}|p{1.4in}|p{0.4in}|} 
\hline
\textbf{Author} & \textbf{Target} & \textbf{Irradiation} & 
\textbf{Beam current measurement and monitor reaction} & \textbf{
Separation method and measurement of activity} & \textbf{Nuclear 
reaction and measured quantity and number of measured data points } & 
\textbf{Covered energy range(MeV)} \\
\hline
\citep{Michel}  & $^{nat}$Zr & cyclotron stacked foil & 
$^{27}$Al(p,x)$^{22}$Na $^{nat}$Cu(p,x)$^{65}$Zn & no 
chemical separation $\gamma$-Ge (Li) & 40-ZR-0(P,X)41-NB-96,SIG, 
23\newline 40-ZR-0(P,X)41-NB-95-M,SIG, 38\newline 40-ZR-0(P,X)41-NB-95-G,SIG, 
52\newline 40-ZR-0(P,X)41-NB-91-M,SIG, 39\newline 40-ZR-0(P,X)41-NB-90-G,M+,SIG, 
72\newline 40-ZR-0(P,X)40-ZR-95,CUM,SIG, 69\newline 40-ZR-0(P,X)40-ZR-89,CUM,SIG, 
91\newline 40-ZR-0(P,X)40-ZR-88,CUM,SIG, 85\newline 40-ZR-0(P,X)40-ZR-86,CUM,SIG, 
30\newline 40-ZR-0(P,X)39-Y-88,SIG, 75\newline 40-ZR-0(P,X)39-Y-87-M,CUM,SIG, 
70\newline 40-ZR-0(P,X)39-Y-87-G,CUM/M-,SIG, 100\newline 40-ZR-0(P,X)39-Y-86-G,M+,SIG, 59 
& 
9.48-43.3\newline 9.48-70.1\newline 9.48-2570\newline 9.48-70.1\newline 9.48-2600\newline 9.48-1600\newline 9.48-2600\newline 18.9-2600\newline 54.4-2600\newline 18.9-2570\newline 9.48-156\newline 9.48-2600\newline 22.2-2570 
\\
\hline
\citep{Busse} & $^{nat}$Zr$^{90}$ZrO$_{2}$ & 
cyclotron stacked foil & $^{63}$Zn(p,n)$^{63}$Zn$^{63}$
Zn(p,2n)$^{62}$Zn & no chemical separation $\gamma$-HPGe & 
40-ZR-0(P,N)41-NB-90,TTY,DT, 5\newline 40-ZR-0(P,N)41-NB-90-G,M+,SIG, 16 \newline
40-ZR-90(P,2N)41-NB-89-G,SIG, 3\newline 40-ZR-90(P,2N)41-NB-89-M,SIG, 
4\newline 40-ZR-90(P,N)41-NB-90,SIG, 12\newline 40-ZR-0(P,N)41-NB-90,TTY,,DT, 5 & 
8.1-17.6\newline 7.5-15.3\newline 17.8-19.0\newline 17.2-19.0\newline 12.0-19.0\newline 11.9-17.6 \\
\hline
\citep{Bringas} & $^{nat}$Zr & cyclotron stacked foil & 
$^{63}$Zn(p,2n)$^{62}$Zn$^{65}$Zn(p,n)$^{65}$Zn & no 
chemical separation $\gamma$-HPGe & 40-ZR-0(P,X)40-ZR-88,CUM,SIG, 2 & 19.6-27.1 
\\
\hline
\citep{Uddin2008} & $^{nat}$Zr & cyclotron stacked foil & $^{
nat}$Cu(p,x)$^{62}$Zn & no chemical separation $\gamma$-HPGe & 
40-ZR-0(P,X)39-Y-86,SIG, 
11\newline 40-ZR-0(P,X)39-Y-87,SIG,15\newline 40-ZR-0(P,X)39-Y-87-M,SIG,12\newline 40-ZR-0(P,X)39-Y-88,SIG,10\newline 40-ZR-0(P,X)40-ZR-88,SIG,9\newline 40-ZR-0(P,X)40-ZR-89,SIG,,15\newline 40-ZR-0(P,X)41-NB-90,SIG, 
15\newline 40-ZR-0(P,X)41-NB-92-M,SIG,15\newline 40-ZR-0(P,X)41-NB-95-G,SIG, 
11\newline 40-ZR-0(P,X)41-NB-96,SIG, 15 & 
18.9-39.7\newline 4.6-39.7\newline 15.1-39.721.5-39.7\newline 24.6-39.7\newline 4.6-39.7\newline 4.6-39.7\newline 4.6-39.7\newline 6.0-39.7\newline 4.6-39.7 
\\
\hline
\citep{Khandaker} & $^{nat}$Zr & cyclotron stacked 
foil & $^{nat}$Cu(p,x)$^{62}$Zn & no chemical separation $\gamma$-HPGe & 
40-ZR-0(P,X)41-NB-92-M,SIG, 17\newline 40-ZR-0(P,X)41-NB-90,SIG, 
16\newline 40-ZR-0(P,X)40-ZR-89-G,CUM,SIG, 16\newline 40-ZR-0(P,X)40-ZR-88,(CUM),SIG, 
8\newline 40-ZR-0(P,X)39-Y-87-M,SIG, 14\newline 40-ZR-0(P,X)39-Y-86,IND,SIG, 10 & 
1.2-40\newline 5.3-40\newline 5.3-40\newline 28-40\newline 11.7-40\newline 24-40 \\
\hline
\citep{Abyad} & $^{nat}$Zr & cyclotron stacked foil 
& $^{nat}$Cu(p,x)$^{65}$Zn & no chemical separation $\gamma$-Ge (Li) & 
40-ZR-0(P,X)41-NB-96,SIG,13\newline 40-ZR-0(P,X)41-NB-95-G,CUM,SIG, 
10\newline 40-ZR-0(P,X)41-NB-92-M,SIG, 13\newline 40-ZR-0(P,X)41-NB-90,SIG, 
11\newline 40-ZR-0(P,X)39-Y-88,SIG, 8 & 4.5-16.7\newline 8.7-16.7\newline 4.5-16.7\newline 7.5-16.7\newline 11-16.9 
\\
\hline
\citep{Naik} & $^{nat}$Zr & cyclotron stacked foil & $^{
nat}$Cu(p,x)$^{62}$Zn & no chemical separation $\gamma$-HPGe & 
40-ZR-90(P,2N)41-NB-89-M/G,SIG/RAT, 1\newline 40-ZR-91(P,3N)41-NB-89-M/G,SIG/RAT, 
1\newline 40-ZR-92(P,4N)41-NB-89-M/G,SIG/RAT, 6 & 19.44\newline 22.58\newline 26.6-44.73 \\
\hline
\citep{Murakami} & $^{nat}$Zr & cyclotron stacked foil 
& $^{nat}$Cu(p,x)$^{62}$Zn & no chemical separation $\gamma$-HPGe & 
40-ZR-0(P,X)41-NB-96,SIG,9\newline 40-ZR-0(P,X)41-NB-95-M,SIG, 
8\newline 40-ZR-0(P,X)41-NB-95-G,SIG, 9\newline 40-ZR-0(P,X)41-NB-92-M,SIG, 
9\newline 40-ZR-0(P,X)41-NB-91-M,SIG, 9\newline 40-ZR-0(P,X)41-NB-90-G,CUM,SIG, 
8\newline 40-ZR-0(P,X)40-ZR-95,SIG, 7\newline 40-ZR-0(P,X)39-Y-87-G,M+,SIG, 
4\newline 40-ZR-0(P,X)39-Y-88,SIG, 8 & 
6.4-14.2\newline 8.6-14.2\newline 6.4-14.2\newline 6.4-14.2\newline 6.4-14.2\newline 7.5-14.2\newline 9.5-14.2\newline 12.0-14.2\newline 8.6-14.2 
\\
\hline
this work & $^{nat}$Zr & cyclotron stacked foil & $^{27}$Al(p,x)
$^{22}$Na $^{nat}$Cu(p,x)$^{65}$Zn$^{nat}$Cu(p,x)$^{62
}$Zn & no chemical separation $\gamma$-HPGe & 40-ZR-0(P,X)41-NB-96,SIG, 19, 40-ZR-0(P,X)41-NB-95-M,SIG, 1240-ZR-0(P,X)41-NB-95-G,SIG, 
31, 40-ZR-0(P,X)41-NB-92-M,SIG,31\newline 40-ZR-0(P,X)41-NB-91-M,SIG, 
12\newline 40-ZR-0(P,X)41-NB-90-G,M+,SIG, 12\newline 40-ZR-0(P,X)40-ZR-95,CUM,SIG, 
30\newline 40-ZR-0(P,X)40-ZR-89,CUM,SIG, 29\newline 40-ZR-0(P,X)40-ZR-88,CUM,SIG, 
29\newline 40-ZR-0(P,X)40-ZR-86,CUM,SIG, 6\newline 40-ZR-0(P,X)39-Y-88,SIG, 
11\newline 40-ZR-0(P,X)39-Y-87-M,CUM,SIG, 30\newline 40-ZR-0(P,X)39-Y-87-G,CUM/M-,SIG, 
30\newline 40-ZR-0(P,X)39-Y-86-G,M+,SIG, 28 & 
20.4-69.8\newline 14.7-36.5\newline 14.7-36.5\newline 14.7-69.8\newline 21.8-49.0\newline 14.7-36.5\newline 17.5-69.8\newline 14.7-69.8\newline 19.6-69.8\newline 53.7-69.8\newline 11.7-36.5\newline 17.5-69.8\newline 14.7-69.8\newline 22.3-69.8 
\\
\hline

\end{tabular}
\begin{flushleft}
\tiny{\noindent SIG-Cross section, TTY-thick target yield, TTD-differential thick target yield, DERIV-derived data, IND-independent formation, CUM-cumulative formation, REL-relative
}
\end{flushleft}

\end{center}
\end{table*}

\subsection{Earlier theoretical estimates and systematics }
\label{2.2}
For estimation of production cross sections of proton induced reaction cross sections a few  systematic theoretical calculations exist  in the TENDL-2013 library \citep{Koning2012} based on TALYS version 1.4 code \citep{Koning2007}, in the MENDL-2p library \citep{Shubin} based on the ALICE-IPPE code \citep{Dityuk}, in the publication of Ren et al. up to 200 MeV \citep{Ren} by using the MEND \citep{Cai} code,  and  in the publication by Sadeghi et  al. and Broeders et al.  based on the ALICE/ASH code \citep{Broeders, Sadeghi}. Nuclear reaction systematics, semi-empirical formulas are also used for the evaluation of reaction cross-section to supplement to the result of measurements and calculations by theoretical models \citep{Tel}. 

\section{Experimental techniques and data evaluation}
\label{3}
\subsection{Experiment}
\label{3.1}
The excitation functions for the $^{nat}$Zr(p,x) reactions were measured at the cyclotrons of the Vrije Universiteit Brussel (VUB, Brussels, Belgium) and of Tohoku University (CYRIC, Sendai, Japan) using the stacked foil technique. The experimental method used was similar to the techniques used in our numerous earlier investigations of charged particle induced nuclear reactions for different applications. Two stacks were irradiated using the 36 MeV (VUB) and 70 MeV (CYRIC) incident proton energy respectively.
In both experiments natural, high purity Zr foils (Goodfellow, $\>$99.98\%, thickness 98.42 $\mu$m and 103 $\mu$m) were assembled together with target foils of other elements for separate investigations and with monitor and degrader foils. 
The target stack at VUB was composed of Ho (26.2 $\mu$m), Zr (98.42 $\mu$m), Al degraders (156.56 $\mu$m), Pb (15.74 $\mu$m) and Ti monitor (12 $\mu$m) foils, repeated 12 times.  The stack composition at higher energy irradiation at CYRIC: Zr(103 $\mu$m), Rh (12.3 $\mu$m), Al (520 $\mu$mm, degrader, and monitor), Mn(10 $\mu$m), Al (520 $\mu$m, degrader, and monitor ) and Ag (52 $\mu$m), repeated 19 times. 
The Ti and Al monitors  foils were used for determination of beam intensity and energy by re-measuring the excitation function for the $^{nat}$Ti(p,x)$^{48}$V reaction at VUB and $^{27}$Al(p,x)$^{22,24}$Na reactions at CYRIC over the entire covered energy range. The target stacks were irradiated in a Faraday-cup like target holder. Irradiations took place at beam current of 160 nA for 60 min (VUB) and 24 nA for 30 min (CYRIC) respectively.
The gamma activity of the majority of the produced radionuclides was measured with standard high purity Ge detectors coupled to acquisition/analysis software. No chemical separation was performed and the measurements were repeated several times up to several months after EOB. Due to large number of simultaneously irradiated targets and the limited detector capacity the first measurements of the induced activity started on both cases after relatively long cooling times  ( VUB:1.5 day, CYRIC: 3 day after EOB).

\subsection{Data processing}
\label{3.2}
For most of the assessed radionuclides different independent $\gamma$-lines are available allowing an internal check of the consistency of the calculated activities. The decay and spectrometric characteristics were taken from the NUDAT2 data base \citep{Nudat} and are summarized in Table 2.
The cross sections were calculated from the well-known activation formula with measured activity, particle flux and number of target nuclei as input parameters. Some of the radionuclides formed are the result of cumulative processes as decay of metastable states or parent nuclides contribute to the production process. The exact physical situation for the individual studied nuclides will be discussed in the next sections.
The particle flux was initially derived through total charge on target by the Faraday cup using a digital integrator. The incident beam energy was determined from the accelerator settings and the mean energy in each foil was calculated by polynomial approximation of Andersen and Ziegler (Andersen and Ziegler, 1977) or calculated with the help of the SRIM code \citep{Ziegler}.
The beam energy and intensity parameters were corrected  by taking into account the comparison of the excitation function of natTi(p,x)48V and $^{27}$Al(p,x)$^{22,24}$Na reactions, re-measured over the whole energy domain studied, with the recommended values in the updated version of IAEA-TECDOC 1211 \citep{TF2001} (Fig. 1). The uncertainty of the incident energy on the first foil in both cases was around $\pm$ 0.3 MeV. Taking into account the cumulative effects of possible variation on incident energy and thickness of the different targets, the uncertainty on the median energy in the last foil was around $\pm$ 1.2 MeV. The uncertainty on each cross-section was estimated in the standard way \citep{error} by taking the square root of the sum in quadrature of all individual contributions, supposing equal sensitivities for the different parameters appearing in the formula. The following individual uncertainties are included in the determination of the peak areas including statistical errors (0.1-20 \%): the number of target nuclei including non-uniformity (5 \%), detector efficiency (5 \%) and incident particle intensity (7 \%). The total uncertainty of the cross-section values was evaluated to vary from 8 to 14 \%, except for few cases where the statistical errors where high.

\begin{figure}
\includegraphics[width=0.5\textwidth]{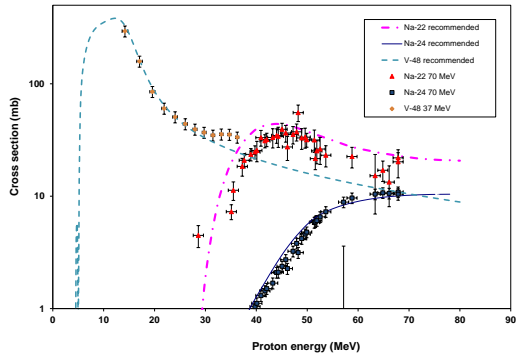}
\caption{The re-measured cross sections of the used monitor reactions $^{27}$Al(p,x)$^{22,24}$Na and  $^{nat}$Ti(p,x)$^{48}$V in comparison with the recommended data}
\label{fig:1}       
\end{figure}

\begin{table*}[t]
\tiny
\caption{Decay characteristics of the investigated activation products and Q-values of contributing reactions}
\begin{center}
\begin{tabular}{|p{1.in}|p{1.in}|p{1.in}|p{1.in}|p{1.in}|p{1.in}|}
\hline
NuclideDecay path & Half-life & E$_{\gamma}$(keV) & I$_{\gamma}$(\%) & 
Contributing reaction & Q-value(keV)GS-GS \\
\hline
\textbf{$^{96}$Nb}\newline $\beta^{-}$: 100 \% & 23.35 h & 
460.040\newline 568.871\newline 778.224\newline 810.330\newline 849.929\newline 1091.349\newline 1200.231 & 
26.62\newline 58.0\newline 96.45\newline 11.09\newline 20.45\newline 48.5\newline 19.97 & $^{96}$Zr(p,n) & -620.13 \\
\hline
\textbf{$^{95m}$Nb}\newline IT: 94.4 $\beta^{-}$: 5.6 \%\newline 235.69\textit{2
} keV & 3.61 d & 235.690 & 24.8 & $^{96}$Zr(p,2n) & -7513.22 \\
\hline
\textbf{$^{95g}$Nb}\newline $\beta^{-}$: 100 \% & 34.991 d & 765.803 & 
99.808 & $^{96}$Zr(p,2n) & -7513.22 \\
\hline
\textbf{$^{92m}$Nb}\newline $\varepsilon$: 100 \%\newline ($\beta^{+}$:0.065 \%)\newline 135.5\textit{4 
}keV & 10.15 d & 934.44 & 99.15 & $^{92}$Zr(p,n)\newline $^{94}$Zr(p,3n)\newline
$^{96}$Zr(p,5n) & -2788.23\newline -17742.15\newline -32058.5 \\
\hline
\textbf{$^{91m}$Nb}\newline IT: 96.6\textit{ }\%$\varepsilon$: 3.4\textit{ }\%$ \beta^
{+}$:0.0013 \%\newline 104.60\textit{5 }keV & 60.86 d & 104.62\newline 1204.67 
& 0.5742.0 & $^{91}$Zr(p,n)\newline $^{92}$Zr(p,2n)\newline $^{94}$Zr(p,4n)\newline $^{
96}$Zr(p,6n) & -2039.93\newline-10674.72\newline -25628.63\newline-39944.98 \\
\hline
\textbf{$^{90}$Nb}\newline $\varepsilon$: 100 \% $\beta^{+}$: 51.2 \% & 14.60 h & 
141.178\newline 1129.224 & 66.8\newline 92.7 & $^{90}$Zr(p,n)\newline $^{91}$Zr(p,2n)\newline
$^{92}$Zr(p,3n)\newline $^{94}$Zr(p,5n)\newline $^{96}$Zr(p,7n) & 
-6893.68\newline -14087.6\newline -22722.41\newline -37676.31\newline -51992.66 \\
\hline
\textbf{$^{95}$Zr}\newline $\beta^{-}$: 100 \% & 64.032 d & 724.192\newline756.725 
& 44.2754.38 & $^{96}$Zr(p,pn)$^{95}\newline$Y decay & -7854.37 \\
\hline
\textbf{$^{89}$Zr}\newline $\varepsilon$: 100 \% $\beta^{+ }$:1.53 \% & 78.41 h & 
909.15 & 99.04 & $^{90}$Zr(p,pn)\newline $^{91}$Zr(p,p2n)\newline $^{92}$
Zr(p,p3n)\newline $^{94}$Zr(p,p5n)\newline $^{96}$Zr(p,p7n)\newline $^{89}$Nb decay & 
-11968.49\newline -19162.4\newline -27797.2\newline -42751.11\newline -57067.45 \\
\hline
\textbf{$^{88}$Zr}\newline $\varepsilon$: 100 \% & 83.4 d & 392.87 & 97.29 & $^{90}$
Zr(p,p2n)\newline $^{91}$Zr(p,p3n)\newline $^{92}$Zr(p,p4n)\newline $^{94}$Zr(p,p6n)\newline
$^{96}$Zr(p,p8n)\newline $^{88}$Nb decay & 
-21287.86\newline-28481.79\newline-37116.59\newline-52070.48\newline-66386.83\newline-29522.6 \\
\hline
\textbf{$^{86}$Zr}\newline $\varepsilon$: 100 \% $\beta^{+ }$:0.05 \% & 16.5 h & 
242.8612.0 & 95.845.8 & $^{90}$Zr(p,p4n)\newline $^{91}$Zr(p,p5n)\newline $^{92
}$Zr(p,p6n)\newline $^{94}$Zr(p,p8n)\newline $^{86}$Nb decay & 
-43090.19\newline-50284.11\newline-58918.9\newline-73872.79\newline-52708.3 \\
\hline
\textbf{$^{88}$Y}\newline$\varepsilon$: 100 \% $\beta^{+}$: 100 \% & 106.627 d & 
898.042\newline 1836.063 &  93.799.2 & $^{90}$Zr(p,2pn)\newline $^{91}$Zr(p,2p2n)\newline
$^{92}$Zr(p,2p3n)\newline $^{94}$Zr(p,2p5n)\newline $^{96}$Zr(p,2p7n) & 
-19835.09\newline-27029.02\newline-35663.81\newline-50617.72\newline-64934.05 \\
\hline
\textbf{$^{87m}$Y}\newline $\varepsilon$: 1.57 \% $\beta^{+ }$:0.75 \% IT: 98.43 
\%\newline 380.82 keV & 13.37 h & 380.79 & 78.06 & $^{90}$Zr(p,2p2n)\newline$^{91
}$Zr(p,2p3n)\newline$^{92}$Zr(p,2p4n)\newline$^{94}$Zr(p,2p6n)\newline$^{96}$
Zr(p,2p8n)\newline$^{87}$Zr decay & 
-29186.82\newline-36380.74\newline-45015.54\newline-59969.44\newline-74285.77\newline-33640.98 \\
\hline
\textbf{$^{87g}$Y}\newline $\varepsilon$: 100 \% $\beta^{+}$:0.180 \% & 79.8 h & 
388.531\newline 484.805 & 82.2\newline89.8 & $^{90}$Zr(p,2p2n)\newline$^{91}$Zr(p,2p3n)\newline
$^{92}$Zr(p,2p4n)\newline$^{94}$Zr(p,2p6n)\newline$^{96}$Zr(p,2p8n)\newline$^{87}$
Zr decay & -29186.82\newline-36380.74\newline-45015.54\newline-59969.44\newline-74285.77\newline-33640.98 \\
\hline
\textbf{$^{86}$Y}\newline $\varepsilon$: 100 \% $\beta^{+}$: 31.9 \% & 14.74 h & 
443.13\newline627.72\newline703.33\newline777.37\newline1076.63\newline1153.05\newline1854.38\newline1920.72 & 
16.9\newline32.6\newline15.4\newline22.4\newline82.5\newline30.5\newline17.2\newline20.8 & $^{90}$Zr(p,2p3n)\newline$^{91}$
Zr(p,2p4n)\newline$^{92}$Zr(p,2p5n)\newline$^{94}$Zr(p,2p7n)\newline$^{96}$
Zr(p,2p9n)\newline$^{86}$Zr decay & 
-40993.3\newline-48187.2\newline-56822.0\newline-71775.9\newline-86092.2\newline-43090.19 \\
\hline
 
\end{tabular}

\begin{flushleft}
\tiny{\noindent When complex particles are emitted instead of individual protons and neutrons the Q-values have to be decreased by the respective binding energies of the compound particles: np-d, +2.2 MeV; 2np-t, +8.48 MeV; 2p2n-$\alpha$, 28.30 MeV. In the case of metastable states a further correction with the level energy in column 1 is also necessary.
}
\end{flushleft}

\end{center}

\end{table*}

\section{RESULTS}
\label{4}
\subsection{Cross sections}
\label{4.1}
The numerical data of excitation functions of $^{96}$Nb, $^{95}$Nb, $^{95g}$Nb, $^{92m}$Nb, $^{91m}$Nb, $^{90}$Nb, $^{95}$Zr, $^{89}$Zr, $^{88}$Zr, $^{86}$Zr, $^{88}$Y, $^{87m}$Y, $^{87g}$Y, $^{86}$Y are presented in Tables 3-5 and are shown in graphical form in Figures 2-15 for comparison with the earlier experimental data and  with the theoretical values taken from the  TENDL-2013 online library \citep{Koning2012} calculated with the 1.4 version of TALYS \citep{Koning2007}. The contributing reactions and decay processes are presented in Table 2. The excitation functions are shortly discussed for each activation product separately.

\begin{table*}[t]
\tiny
\caption{Activation cross sections of niobium radioisotopes in proton induced reactions on zirconium}
\begin{center}
\begin{tabular}{|l|l|l|l|l|l|l|l|l|l|l|l|l|l|l|l|l|l|l|l|l|}
\hline
\multicolumn{3}{|c|}{\textbf{Energy(MeV)}} & \multicolumn{3}{|c|}{\textbf{$^{96}$Nb(mbarn)}} & \multicolumn{3}{|c|}{\textbf{$^{
95m}$Nb (mbarn)}} & \multicolumn{3}{|c|}{\textbf{$^{95g}$Nb (mbarn)}} & \multicolumn{3}{|c|}{\textbf{$^{
92m}$Nb (mbarn)}} & \multicolumn{3}{|c|}{\textbf{$^{91m}$Nb (mbarn)}} & \multicolumn{3}{|c|}{\textbf{$^{90
}$Nb (mbarn)}} \\
\hline
\multicolumn{21}{|c|}{\textbf{VUB ser. 1}} \\
\hline
36.5 & $\pm$ & 0.3 & & & & 0.44 & $\pm$ & 0.09 & 2.56 & $\pm$ & 0.29 & 20.1 & $\pm$ & 
2.2 & & & & 102.6 & $\pm$ & 11.1 \\
\hline
35.0 & $\pm$ & 0.3 & & & & 0.68 & $\pm$ & 0.13 & 2.71 & $\pm$ & 0.31 & 24.2 & $\pm$ & 
2.6 & & & & 107.8 & $\pm$ & 11.7 \\
\hline
33.4 & $\pm$ & 0.4 & & & & 0.71 & $\pm$ & 0.09 & 2.71 & $\pm$ & 0.30 & 30.3 & $\pm$ & 
3.3 & & & & 112.3 & $\pm$ & 12.2 \\
\hline
31.8 & $\pm$ & 0.4 & & & & 0.57 & $\pm$ & 0.12 & 2.88 & $\pm$ & 0.33 & 38.1 & $\pm$ & 
4.1 & & & & 117.0 & $\pm$ & 12.7 \\
\hline
30.1 & $\pm$ & 0.4 & & & & 0.78 & $\pm$ & 0.16 & 3.04 & $\pm$ & 0.35 & 46.0 & $\pm$ & 
5.0 & & & & 117.8 & $\pm$ & 12.8 \\
\hline
28.3 & $\pm$ & 0.5 & & & & 0.76 & $\pm$ & 0.17 & 3.00 & $\pm$ & 0.35 & 52.5 & $\pm$ & 
5.7 & & & & 116.4 & $\pm$ & 12.6 \\
\hline
26.4 & $\pm$ & 0.5 & & & & 0.87 & $\pm$ & 0.16 & 3.96 & $\pm$ & 0.44 & 54.8 & $\pm$ & 
5.9 & & & & 114.5 & $\pm$ & 12.4 \\
\hline
24.4 & $\pm$ & 0.6 & & & & 1.02 & $\pm$ & 0.12 & 4.91 & $\pm$ & 0.53 & 53.8 & $\pm$ & 
5.8 & 21.1 & $\pm$ & 4.0 & 107.2 & $\pm$ & 11.6 \\
\hline
22.3 & $\pm$ & 0.6 & & & & 1.55 & $\pm$ & 0.22 & 7.26 & $\pm$ & 0.80 & 45.5 & $\pm$ & 
4.9 & & & & 113.6 & $\pm$ & 12.3 \\
\hline
20.0 & $\pm$ & 0.7 & & & & 2.70 & $\pm$ & 0.31 & 12.32 & $\pm$ & 1.33 & 23.1 & $\pm$ & 
2.5 & & & & 150.1 & $\pm$ & 16.3 \\
\hline
17.5 & $\pm$ & 0.7 & & & & 4.60 & $\pm$ & 0.51 & 20.84 & $\pm$ & 2.26 & 5.4 & $\pm$ & 
0.6 & 21.8 & $\pm$ & 5.5 & 240.6 & $\pm$ & 26.1 \\
\hline
14.7 & $\pm$ & 0.8 & & & & 6.00 & $\pm$ & 1.00 & 20.82 & $\pm$ & 2.92 & 10.0 & $\pm$ & 
2.0 &   & & & 423.9 & $\pm$ & 45.9 \\
\hline
\multicolumn{21}{|c|}{\textbf{CYRIC, ser. 2}} \\
\hline
69.8 & $\pm$ & 0.3 & 1.73 & $\pm$ & 0.44 & & & & 1.35 & $\pm$ & 0.29 & 6.84 & $\pm$ & 
0.75 & & & & & & \\
\hline
66.7 & $\pm$ & 0.4 & 1.93 & $\pm$ & 0.66 & & & & 1.39 & $\pm$ & 0.27 & 7.47 & $\pm$ & 
0.82 & & & & & & \\
\hline
63.9 & $\pm$ & 0.4 & 1.87 & $\pm$ & 0.63 & & & & 1.31 & $\pm$ & 0.30 & 7.89 & $\pm$ & 
0.87 & & & & & & \\
\hline
60.5 & $\pm$ & 0.5 & 1.98 & $\pm$ & 0.86 & & & & 1.56 & $\pm$ & 0.31 & 8.64 & $\pm$ & 
0.95 & & & & & & \\
\hline
57.4 & $\pm$ & 0.6 & 1.51 & $\pm$ & 0.60 & & & & 1.53 & $\pm$ & 0.31 & 9.36 & $\pm$ & 
1.03 & & & & & & \\
\hline
53.7 & $\pm$ & 0.7 & 1.68 & $\pm$ & 0.73 & & & & 1.98 & $\pm$ & 0.35 & 10.3 & $\pm$ & 
1.1 & & & & & & \\
\hline
51.2 & $\pm$ & 0.7 & 1.64 & $\pm$ & 0.77 & & & & 2.02 & $\pm$ & 0.39 & 11.1 & $\pm$ & 
1.2 & & & & & & \\
\hline
49.0 & $\pm$ & 0.8 & 1.09 & $\pm$ & 0.45 & & & & 1.76 & $\pm$ & 0.34 & 11.8 & $\pm$ & 
1.3 & 15.5 & $\pm$ & 6.6 & & & \\
\hline
47.2 & $\pm$ & 0.8 & 1.37 & $\pm$ & 0.73 & & & & 1.63 & $\pm$ & 0.36 & 12.1 & $\pm$ & 
1.3 & & & & & & \\
\hline
44.9 & $\pm$ & 0.8 & 1.36 & $\pm$ & 0.51 & & & & 2.19 & $\pm$ & 0.38 & 13.0 & $\pm$ & 
1.4 & 25.2 & $\pm$ & 8.6 & & & \\
\hline
43.0 & $\pm$ & 0.9 & 1.07 & $\pm$ & 0.44 & & & & 2.19 & $\pm$ & 0.37 & 12.7 & $\pm$ & 
1.4 & 19.5 & $\pm$ & 7.0 & & & \\
\hline
40.5 & $\pm$ & 0.9 & 2.24 & $\pm$ & 0.84 & & & & 1.98 & $\pm$ & 0.34 & 13.7 & $\pm$ & 
1.5 & 14.8 & $\pm$ & 7.7 & & & \\
\hline
38.4 & $\pm$ & 1.0 & 2.11 & $\pm$ & 0.97 & & & & 2.38 & $\pm$ & 0.39 & 16.0 & $\pm$ & 
1.7 & 14.4 & $\pm$ & 6.6 & & & \\
\hline
35.7 & $\pm$ & 1.1 & 2.36 & $\pm$ & 0.89 & & & & 2.53 & $\pm$ & 0.43 & 21.9 & $\pm$ & 
2.4 & 17.5 & $\pm$ & 6.7 & & & \\
\hline
33.3 & $\pm$ & 1.1 & 1.80 & $\pm$ & 0.64 & & & & 2.23 & $\pm$ & 0.40 & 30.8 & $\pm$ & 
3.3 & 10.8 & $\pm$ & 6.3 & & & \\
\hline
30.2 & $\pm$ & 1.2 & 2.52 & $\pm$ & 1.16 & & & & 3.20 & $\pm$ & 0.52 & 45.6 & $\pm$ & 
4.9 & & & & & & \\
\hline
27.4 & $\pm$ & 1.2 & 2.37 & $\pm$ & 1.43 & & & & 3.06 & $\pm$ & 0.55 & 53.5 & $\pm$ & 
5.8 & 20.2 & $\pm$ & 9.3 & & & \\
\hline
23.8 & $\pm$ & 1.3 & 1.51 & $\pm$ & 0.89 & & & & 5.80 & $\pm$ & 0.74 & 49.3 & $\pm$ & 
5.3 & 17.8 & $\pm$ & 8.4 & & & \\
\hline
20.4 & $\pm$ & 1.4 & 2.35 & $\pm$ & 0.99 & & & & 11.94 & $\pm$ & 1.30 & 25.8 & $\pm$ & 
2.8 & 49.5 & $\pm$ & 6.0 & & & \\
\hline
\end{tabular}

\end{center}
\end{table*}








\begin{table*}[t]
\tiny
\caption{Activation cross sections of zirconium radioisotopes in proton induced reactions on zirconium}
\begin{center}
\begin{tabular}{|l|l|l|l|l|l|l|l|l|l|l|l|l|l|l|}
\hline
\multicolumn{3}{|c|}{\textbf{Energy(MeV)}} & \multicolumn{3}{|c|}{\textbf{$^{95}$Zr(mbarn)}} & \multicolumn{3}{|c|}{\textbf{$^{
89g}$Zr (mbarn)}} & \multicolumn{3}{|c|}{\textbf{$^{88}$Zr (mbarn)}} & \multicolumn{3}{|c|}{\textbf{$^{86
}$Zr (mbarn)}} \\
\hline
\multicolumn{15}{|c|}{\textbf{VUB ser. 1}} \\
\hline
36.5 & $\pm$ & 0.3 & 4.25 & $\pm$ & 0.51 & 274.3 & $\pm$ & 29.7 & 237.4 & $\pm$ & 25.9 & & & \\
\hline
35.0 & $\pm$ & 0.3 & 4.57 & $\pm$ & 0.58 & 329.8 & $\pm$ & 35.7 & 220.0 & $\pm$ & 23.9 & & & \\
\hline
33.4 & $\pm$ & 0.4 & 4.13 & $\pm$ & 0.46 & 347.6 & $\pm$ & 37.6 & 169.0 & $\pm$ & 18.6 & & & \\
\hline
31.8 & $\pm$ & 0.4 & 3.72 & $\pm$ & 0.49 & 428.6 & $\pm$ & 46.4 & 121.5 & $\pm$ & 13.3 & & & \\
\hline
30.1 & $\pm$ & 0.4 & 3.84 & $\pm$ & 0.51 & 420.9 & $\pm$ & 45.5 & 64.7 & $\pm$ & 7.3 & & & \\
\hline
28.3 & $\pm$ & 0.5 & 3.29 & $\pm$ & 0.49 & 437.7 & $\pm$ & 47.4 & 29.4 & $\pm$ & 3.9 & & & \\
\hline
26.4 & $\pm$ & 0.5 & 3.78 & $\pm$ & 0.55 & 451.7 & $\pm$ & 48.9 & & & & & & \\
\hline
24.4 & $\pm$ & 0.6 & 4.14 & $\pm$ & 0.47 & 426.3 & $\pm$ & 46.1 & & & & & & \\
\hline
22.3 & $\pm$ & 0.6 & 3.59 & $\pm$ & 0.51 & 399.0 & $\pm$ & 43.2 & & & & & & \\
\hline
20.0 & $\pm$ & 0.7 & 3.80 & $\pm$ & 0.48 & 340.0 & $\pm$ & 36.8 & & & & & & \\
\hline
17.5 & $\pm$ & 0.7 & 2.42 & $\pm$ & 0.38 & 194.7 & $\pm$ & 21.1 & & & & & & \\
\hline
14.7 & $\pm$ & 0.8 & & & & 27.1 & $\pm$ & 3.2 & & & & & & \\
\hline
\multicolumn{15}{|c|}{\textbf{CYRIC, ser. 2}} \\
\hline
69.8 & $\pm$ & 0.3 & 3.61 & $\pm$ & 1.17 & 199.2 & $\pm$ & 21.6 & 174.4 & $\pm$ & 18.9 & 44.3 & $\pm$ & 4.9 \\
\hline
66.7 & $\pm$ & 0.4 & 3.31 & $\pm$ & 1.09 & 203.5 & $\pm$ & 22.0 & 185.8 & $\pm$ & 20.1 & 35.2 & $\pm$ & 3.9 \\
\hline
63.9 & $\pm$ & 0.4 & 2.93 & $\pm$ & 1.23 & 205.8 & $\pm$ & 22.3 & 190.2 & $\pm$ & 20.6 & 25.3 & $\pm$ & 2.9 \\
\hline
60.5 & $\pm$ & 0.5 & 3.51 & $\pm$ & 1.36 & 211.0 & $\pm$ & 22.8 & 197.9 & $\pm$ & 21.4 & 13.3 & $\pm$ & 1.7 \\
\hline
57.4 & $\pm$ & 0.6 & 3.50 & $\pm$ & 1.25 & 218.0 & $\pm$ & 23.6 & 202.1 & $\pm$ & 21.9 & 5.05 & $\pm$ & 0.92 \\
\hline
53.7 & $\pm$ & 0.7 & 3.82 & $\pm$ & 1.31 & 235.8 & $\pm$ & 25.5 & 218.6 & $\pm$ & 23.7 & 1.02 & $\pm$ & 0.44 \\
\hline
51.2 & $\pm$ & 0.7 & 3.19 & $\pm$ & 1.35 & 247.7 & $\pm$ & 26.8 & 238.8 & $\pm$ & 25.9 & & & \\
\hline
49.0 & $\pm$ & 0.8 & 4.18 & $\pm$ & 1.30 & 255.0 & $\pm$ & 27.6 & 263.2 & $\pm$ & 28.5 & & & \\
\hline
47.2 & $\pm$ & 0.8 & 3.68 & $\pm$ & 1.35 & 255.9 & $\pm$ & 27.7 & 285.0 & $\pm$ & 30.9 & & & \\
\hline
44.9 & $\pm$ & 0.8 & 4.09 & $\pm$ & 1.25 & 276.8 & $\pm$ & 30.0 & 327.5 & $\pm$ & 35.5 & & & \\
\hline
43.0 & $\pm$ & 0.9 & 4.00 & $\pm$ & 1.57 & 262.8 & $\pm$ & 28.4 & 317.3 & $\pm$ & 34.4 & & & \\
\hline
40.5 & $\pm$ & 0.9 & 4.88 & $\pm$ & 1.24 & 266.1 & $\pm$ & 28.8 & 312.8 & $\pm$ & 33.9 & & & \\
\hline
38.4 & $\pm$ & 1.0 & 3.58 & $\pm$ & 1.36 & 272.2 & $\pm$ & 29.5 & 291.5 & $\pm$ & 31.6 & & & \\
\hline
35.7 & $\pm$ & 1.1 & 3.57 & $\pm$ & 1.33 & 305.5 & $\pm$ & 33.1 & 247.5 & $\pm$ & 26.8 & & & \\
\hline
33.3 & $\pm$ & 1.1 & 3.38 & $\pm$ & 1.34 & 361.4 & $\pm$ & 39.1 & 179.1 & $\pm$ & 19.4 & & & \\
\hline
30.2 & $\pm$ & 1.2 & 4.82 & $\pm$ & 1.98 & 448.4 & $\pm$ & 48.5 & 75.6 & $\pm$ & 8.2 & & & \\
\hline
27.4 & $\pm$ & 1.2 & 4.26 & $\pm$ & 2.20 & 469.4 & $\pm$ & 50.8 & 17.1 & $\pm$ & 2.1 & & & \\
\hline
23.8 & $\pm$ & 1.3 & 3.95 & $\pm$ & 1.66 & 438.4 & $\pm$ & 47.4 & & & & & & \\
\hline
20.4 & $\pm$ & 1.4 & 3.55 & $\pm$ & 0.57 & 360.7 & $\pm$ & 39.0 & & & & & & \\
\hline
\end{tabular}
\end{center}
\end{table*}

\begin{table*}[t]
\tiny
\caption{Activation cross sections of yttrium radioisotopes in proton induced reactions on zirconium}
\begin{center}
\begin{tabular}{|l|l|l|l|l|l|l|l|l|l|l|l|l|l|l|}
\hline
\multicolumn{3}{|c|}{\textbf{Energy(MeV)}} & \multicolumn{3}{|c|}{\textbf{$^{88}$Y(mbarn)}} & \multicolumn{3}{|c|}{\textbf{$^{
87m}$Y(mbarn)}} & \multicolumn{3}{|c|}{\textbf{$^{87g}$Y (mbarn)}} & \multicolumn{3}{|c|}{\textbf{$^{86}$Y (mbarn)}} \\
\hline
\multicolumn{15}{|c|}{\textbf{VUB ser. 1}} \\
\hline
36.5 & $\pm$ & 0.3 & 19.8 & $\pm$ & 4.3 & 8.44 & $\pm$ & 0.94 & 11.8 & $\pm$ & 1.3 & 23.3 & $\pm$ & 2.6 \\
\hline
35.0 & $\pm$ & 0.3 & 17.0 & $\pm$ & 3.7 & 8.31 & $\pm$ & 0.91 & 11.6 & $\pm$ & 1.3 & 22.7 & $\pm$ & 2.5 \\
\hline
33.4 & $\pm$ & 0.4 & 16.3 & $\pm$ & 3.5 & 8.01 & $\pm$ & 0.89 & 10.9 & $\pm$ & 1.2 & 21.3 & $\pm$ & 2.3 \\
\hline
31.8 & $\pm$ & 0.4 & 13.9 & $\pm$ & 3.0 & 7.26 & $\pm$ & 0.80 & 10.7 & $\pm$ & 1.2 & 18.5 & $\pm$ & 2.0 \\
\hline
30.1 & $\pm$ & 0.4 & 10.0 & $\pm$ & 2.2 & 6.48 & $\pm$ & 0.73 & 9.01 & $\pm$ & 1.00 & 15.8 & $\pm$ & 1.7 \\
\hline
28.3 & $\pm$ & 0.5 & 6.63 & $\pm$ & 1.46 & 6.43 & $\pm$ & 0.73 & 9.02 & $\pm$ & 1.00 & 13.2 & $\pm$ & 1.5 \\
\hline
26.4 & $\pm$ & 0.5 & 5.94 & $\pm$ & 0.68 & 6.92 & $\pm$ & 0.79 & 10.4 & $\pm$ & 1.2 & 9.44 & $\pm$ & 1.06 \\
\hline
24.4 & $\pm$ & 0.6 & 4.56 & $\pm$ & 0.50 & 7.07 & $\pm$ & 0.80 & 11.6 & $\pm$ & 1.3 & 4.39 & $\pm$ & 0.52 \\
\hline
22.3 & $\pm$ & 0.6 & 3.33 & $\pm$ & 0.43 & 7.84 & $\pm$ & 0.88 & 11.7 & $\pm$ & 1.3 & 0.80 & $\pm$ & 0.16 \\
\hline
20.0 & $\pm$ & 0.7 & 2.57 & $\pm$ & 0.32 & 5.11 & $\pm$ & 0.59 & 9.76 & $\pm$ & 1.09 & & & \\
\hline
17.5 & $\pm$ & 0.7 & 1.89 & $\pm$ & 0.26 & 3.02 & $\pm$ & 0.38 & 5.48 & $\pm$ & 0.64 & & & \\
\hline
14.7 & $\pm$ & 0.8 & & & & & & & & & & & & \\
\hline
\multicolumn{3}{|c|}{\textbf{CYRIC, ser. 2}} \\
\hline
69.8 & $\pm$ & 0.3 & & & & 203.0 & $\pm$ & 22.2 & 221.8 & $\pm$ & 24.0 & 225.2 & $\pm$ & 24.5 \\
\hline
66.7 & $\pm$ & 0.4 & & & & 205.9 & $\pm$ & 22.5 & 235.9 & $\pm$ & 25.5 & 180.1 & $\pm$ & 19.7 \\
\hline
63.9 & $\pm$ & 0.4 & & & & 214.7 & $\pm$ & 23.5 & 248.4 & $\pm$ & 26.9 & 129.1 & $\pm$ & 14.2 \\
\hline
60.5 & $\pm$ & 0.5 & & & & 211.2 & $\pm$ & 23.0 & 259.9 & $\pm$ & 28.1 & 75.1 & $\pm$ & 8.3 \\
\hline
57.4 & $\pm$ & 0.6 & & & & 209.6 & $\pm$ & 22.8 & 250.4 & $\pm$ & 27.1 & 73.5 & $\pm$ & 8.2 \\
\hline
53.7 & $\pm$ & 0.7 & & & & 185.4 & $\pm$ & 20.1 & 224.1 & $\pm$ & 24.2 & 19.3 & $\pm$ & 2.2 \\
\hline
51.2 & $\pm$ & 0.7 & & & & 153.8 & $\pm$ & 16.7 & 186.6 & $\pm$ & 20.2 & 16.5 & $\pm$ & 1.9 \\
\hline
49.0 & $\pm$ & 0.8 & & & & 122.2 & $\pm$ & 13.4 & 146.8 & $\pm$ & 15.9 & 14.9 & $\pm$ & 1.8 \\
\hline
47.2 & $\pm$ & 0.8 & & & & 87.1 & $\pm$ & 9.6 & 108.4 & $\pm$ & 11.7 & 14.8 & $\pm$ & 1.8 \\
\hline
44.9 & $\pm$ & 0.8 & & & & 52.1 & $\pm$ & 5.7 & 67.5 & $\pm$ & 7.3 & 15.3 & $\pm$ & 1.7 \\
\hline
43.0 & $\pm$ & 0.9 & & & & 26.0 & $\pm$ & 3.4 & 35.5 & $\pm$ & 3.9 & 17.0 & $\pm$ & 2.1 \\
\hline
40.5 & $\pm$ & 0.9 & & & & 15.3 & $\pm$ & 2.3 & 19.1 & $\pm$ & 2.1 & 20.2 & $\pm$ & 2.4 \\
\hline
38.4 & $\pm$ & 1.0 & & & & 10.6 & $\pm$ & 2.1 & 15.5 & $\pm$ & 1.7 & 22.6 & $\pm$ & 2.6 \\
\hline
35.7 & $\pm$ & 1.1 & & & & 10.7 & $\pm$ & 1.9 & 14.0 & $\pm$ & 1.5 & 25.2 & $\pm$ & 2.9 \\
\hline
33.2 & $\pm$ & 1.1 & & & & 7.99 & $\pm$ & 1.82 & 12.9 & $\pm$ & 1.4 & 21.3 & $\pm$ & 2.5 \\
\hline
30.2 & $\pm$ & 1.2 & & & & & & & 11.2 & $\pm$ & 1.2 & 18.8 & $\pm$ & 2.2 \\
\hline
27.4 & $\pm$ & 1.2 & & & & 4.19 & $\pm$ & 1.66 & 11.3 & $\pm$ & 1.2 & 11.1 & $\pm$ & 1.4 \\
\hline
23.8 & $\pm$ & 1.3 & & & & 6.84 & $\pm$ & 2.43 & 13.8 & $\pm$ & 1.5 & 2.55 & $\pm$ & 0.74 \\
\hline
20.4 & $\pm$ & 1.4 & & & & 5.44 & $\pm$ & 2.03 & 11.9 & $\pm$ & 1.3 & & & \\
\hline
\end{tabular}
\end{center}
\end{table*}

\subsubsection{Cross sections of $^{96}$Nb}
\label{4.1.1}
The radionuclide $^{96}$Nb (T$_{1/2}$ = 23.35 h) can only be produced via the $^{96}$Zr(p,n) reaction. Due to the experimental circumstances we could obtain data only from the high energy irradiation (Fig. 2) and the results have large uncertainties due to the low counting statistics. The agreement with the literature data and with the theory is acceptable.

\begin{figure}
\includegraphics[width=0.5\textwidth]{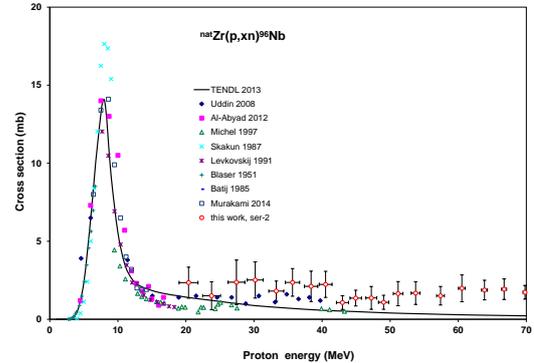}
\caption{Excitation function of the $^{nat}$Zr(p,x)$^{96}$Nb reaction}
\label{fig:2}       
\end{figure}

\subsubsection{Cross sections of $^{95}$Nb}
\label{4.1.2}
The radionuclide $^{95}$Nb has shorter-lived metastable state $^{95}$Nb (T$_{1/2}$ = 86.64 h) and a longer-lived ground state $^{95g}$Nb (T$_{1/2}$ = 34.991 d). The isomeric state decays to the ground state by 94.4\% IT process. We obtained experimental data for $^{95}$Nb only from the low energy irradiation, the 235 keV $\gamma$-line of the metastable state could not be separated reliably in our spectra from the high energy experiment. Our results and the data of (Levkovskii, 1991) are in good agreement (Fig. 3). The data of \citep{Michel} are surprisingly higher by a factor of five. The TENDL-2013 follows the trend of the experimental data but gives lower values.

\begin{figure}
\includegraphics[width=0.5\textwidth]{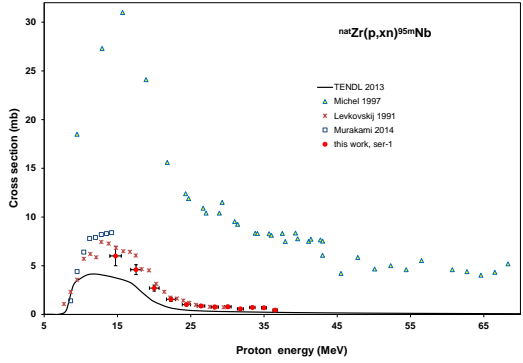}
\caption{Excitation function of the $^{nat}$Zr(p,x)$^{95}$Nb reaction}
\label{fig:3}       
\end{figure}

\subsubsection{Cross sections of $^{95g}$Nb}
\label{4.1.3}

The measured cross-sections of the $^{95g}$Nb (T$_{1/2}$ = 34.991 d) are cumulative, as they are deduced from spectra taken after a cooling time resulting in nearly complete  IT decay of $^{95g}$Nb (T$_{1/2}$ = 86.64 h). The agreement with the earlier experimental data is acceptable for both experiments (Fig. 4). The results of the nuclear model code TALYS from the TENDL-2013 library are in good agreement with the experiments.

\begin{figure}
\includegraphics[width=0.5\textwidth]{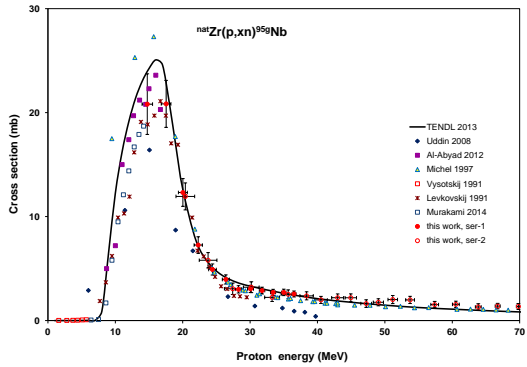}
\caption{Excitation function of the $^{nat}$Zr(p,x)$^{95g}$Nb reaction}
\label{fig:4}       
\end{figure}

\subsubsection{Cross sections of $^{92m}$Nb}
\label{4.1.4}
The half-life of the ground state of the $^{92}$Nb is very long (T$_{1/2}$ = 3.47 107 a) and 92gNb can hence be considered as stable in our experimental conditions. The isomeric state $^{92m}$Nb (T$_{1/2}$ = 10.15 d) has no IT, but decays directly to stable 92Zr. Our results for both experiments are in acceptable agreement with the earlier experimental results and with the data in TENDL-2013 (Fig. 5).

\begin{figure}
\includegraphics[width=0.5\textwidth]{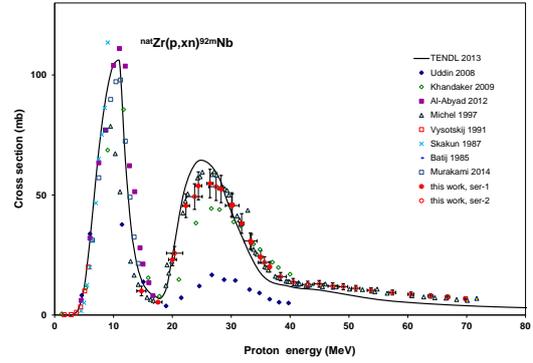}
\caption{Excitation function of the $^{nat}$Zr(p,x)$^{92m}$Nb reaction}
\label{fig:5}       
\end{figure}

\subsubsection{Cross sections of $^{91m}$Nb}
\label{4.1.5}
The very long-lived ground state of $^{91}$Nb (T$_{1/2}$ = 6800 a) has no measurable gamma lines. We obtained some results for the activation cross sections of the shorter-lived isomeric state (T$_{1/2}$ = 60.86 d), mostly from the high energy irradiation. Our data show large uncertainties due to the low statistics. The agreement with the earlier experimental results is acceptable (Fig. 6). The TENDL-2013 prediction overestimates the experimental data.

\begin{figure}
\includegraphics[width=0.5\textwidth]{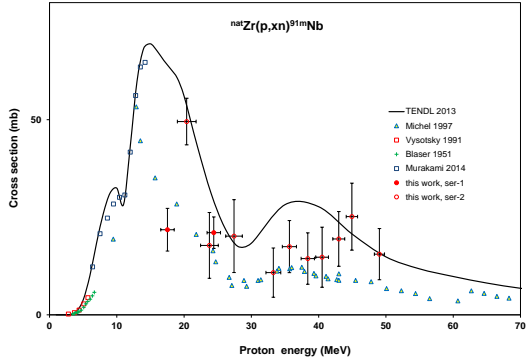}
\caption{Excitation function of the $^{nat}$Zr(p,x)$^{91m}$Nb reaction}
\label{fig:6}       
\end{figure}

\subsubsection{Cross sections of $^{90}$Nb}
\label{4.1.6}
The cross-sections for production of $^{90g}$Nb (T$_{1/2}$ = 14.6 h) obtained in the VUB experiment are shown in Fig. 7. The cross sections include, apart from the direct production, the contribution through IT decay (100\%) of the 18.8 s half-life isomeric state. The agreement with the earlier experimental results is good, except for the data of \citep{Kondratev, Blosser, Birjukov, Abyad} (Fig. 7). The TENDL-2013 results are well representing the energy dependence of the contributions of the different reactions and their maximum values. 

\begin{figure}
\includegraphics[width=0.5\textwidth]{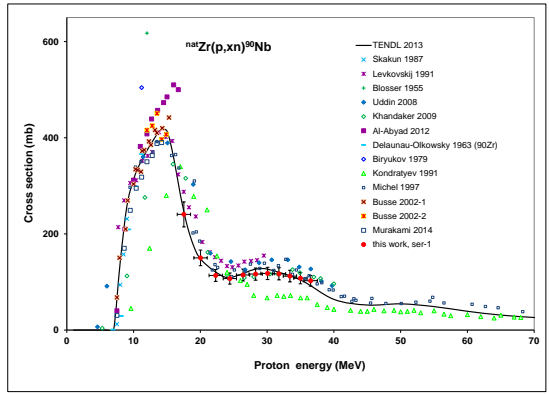}
\caption{Excitation function of the $^{nat}$Zr(p,x)$^{90}$Nb reaction}
\label{fig:7}       
\end{figure}

\subsubsection{Cross sections of $^{95}$Zr}
\label{4.1.7}
The radionuclide $^{95}$Zr (T$_{1/2}$ = 64.032 d) is formed directly through the $^{96}$Zr(p,pn) reaction and indirectly from the decay of  the short-lived parent $^{95}$Y (T$_{1/2}$ = 10.3 min) produced via the $^{96}$Zr(p,2p) reaction. The agreement with earlier experimental data and with the TENDL-2013 predictions is acceptable for our 2 new experiments (Fig. 8). 

\begin{figure}
\includegraphics[width=0.5\textwidth]{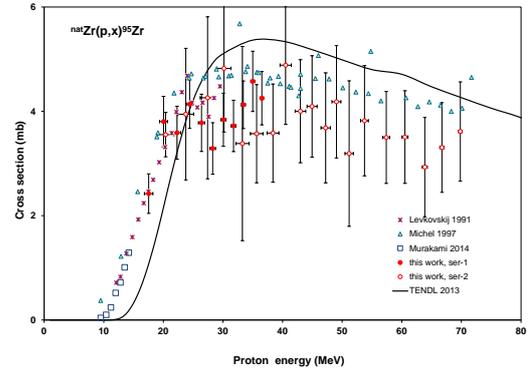}
\caption{Excitation function of the $^{nat}$Zr(p,x)$^{95}$Zr reaction}
\label{fig:8}       
\end{figure}

\subsubsection{Cross sections of $^{89}$Zr}
\label{4.1.8}
The cross sections for $^{89g}$Zr (T$_{1/2}$ = 78.41 h) production include, apart from the direct (p,pxn) processes on different stable Zr isotopes, also the contributions from the decay of the short-lived metastable parent $^{89m}$Zr (T$_{1/2}$ = 4.161 min) and the decay of $^{89}$Nb (T$_{1/2}$ = 66 min). Our data from both experiments do not support the low cross section values of \citep{Khandaker} and the TENDL-2013 predictions near in the 20-30 MeV energy range but are in agreement with the other literature values (Fig. 9).

\begin{figure}
\includegraphics[width=0.5\textwidth]{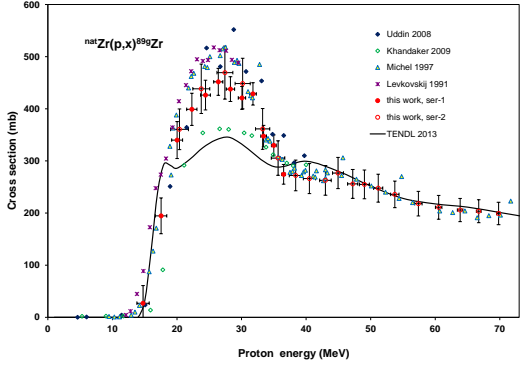}
\caption{Excitation function of the $^{nat}$Zr(p,x)$^{89}$Zr reaction}
\label{fig:9}       
\end{figure}

\subsubsection{Cross sections of $^{88}$Zr}
\label{4.1.9}
The measured cross-sections for $^{88}$Zr (T$_{1/2}$ = 83.4 d) production contain the direct formation and contributions from the decay of the two states of the short-lived parent $^{88m,g}$Nb (T$_{1/2}$ =7.78 min and 14.55 min). Our both experimental data sets support the high maximum as found by \citep{Michel}. The agreement with the TENDL-2013 is acceptable (Fig. 10).
\begin{figure}
\includegraphics[width=0.5\textwidth]{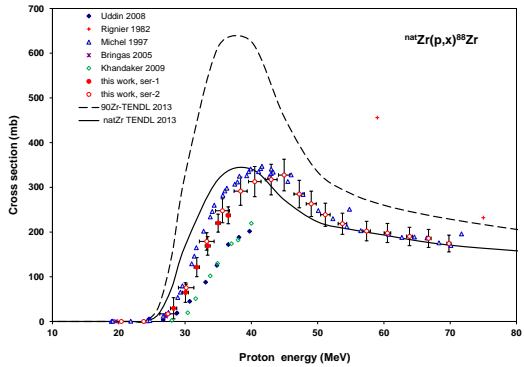}
\caption{Excitation function of the $^{nat}$Zr(p,x)$^{88}$Zr reaction}
\label{fig:10}       
\end{figure}

\subsubsection{Cross sections of $^{86}$Zr}
\label{4.1.10}
The radionuclide $^{86}$Zr (16.5 h) was produced directly and from the decay of the two states of $^{86m,g}$Nb (T$_{1/2}$ = 56.3 s and 88 s).Our data agree with the experimental data of \citep{Michel} and are lower than the TENDL-2013 results (Fig. 11).

\begin{figure}
\includegraphics[width=0.5\textwidth]{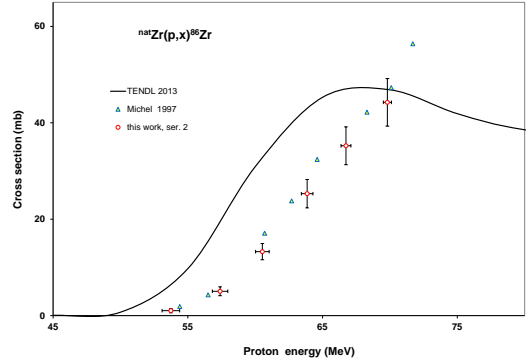}
\caption{Excitation function of the $^{nat}$Zr(p,x)$^{86}$Zr reaction}
\label{fig:11}       
\end{figure}

\subsubsection{Cross sections of $^{88}$Y}
\label{4.1.11}
The cross sections of $^{88}$Y (T$_{1/2}$ = 106.627 d) were obtained from the first spectra measured after EOB in order to minimize the contribution from the decay of the $^{88}$Zr parent nuclide with similar half-life (T$_{1/2}$ = 83.4 d).  Up to 22-25MeV no contribution from decay (also not from parent $^{88}$Nb). Decay contribution does exist at higher energies. To get independent cross sections, the contribution from the decay of $^{88}$Zr was subtracted based on the measured counts at 393 keV, resulting in corrections on the count rate of the independent  898 keV $\gamma$-line. Unfortunately, because of the relative long half-life and the shorter measuring time the $^{88}$Y peak could not be evaluated reliably from the high energy spectra.
 The agreement is acceptable (Fig. 12).

\begin{figure}
\includegraphics[width=0.5\textwidth]{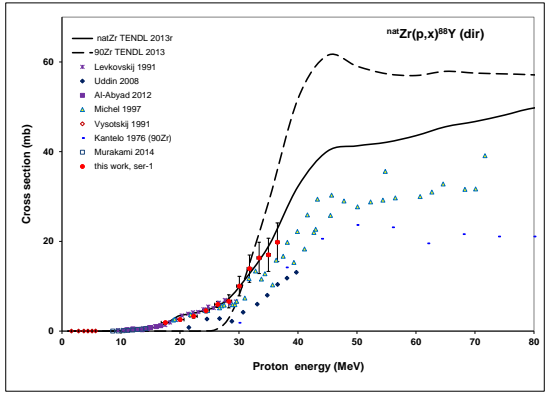}
\caption{Excitation function of the $^{nat}$Zr(p,x)$^{88}$Y reaction}
\label{fig:12}       
\end{figure}

\subsubsection{Cross sections of $^{87m}$Y}
\label{4.1.12}
The radionuclide $^{87}$Y has a metastable state $^{87m}$Y (T$_{1/2}$ = 13.37 h) that decays completely to the ground state $^{87g}$Y (T$_{1/2}$ = 79.8 h). The cross-sections for $^{87m}$Y (Fig.13) contain the contributions from both the direct production through (p,2pxn) reactions and the decay of $^{87g}$Zr (T$_{1/2}$ = 1.68 h, $\varepsilon$ = 100 \%) that could not be assessed independently. As the reaction with lowest threshold to produce $^{87}$Zr is the $^{90}$Zr(p,p3n) process with a threshold above 33 MeV, the cross sections at low energy are due to emission of clusters ($\alpha$-particles) in the direct reaction. The results of our two new data sets agree well with the earlier studies both in the low (cluster emission) and high energy (individual nucleons) regions. The TENDL-2013 predictions underestimate by a factor of 6 the high energy individual nucleons emission.

\begin{figure}
\includegraphics[width=0.5\textwidth]{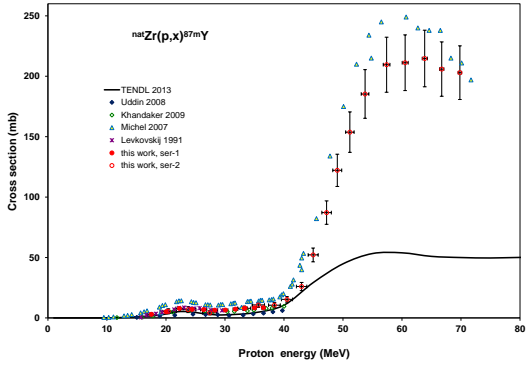}
\caption{Excitation function of the $^{nat}$Zr(p,x)$^{87m}$Y reaction}
\label{fig:13}       
\end{figure}

\subsubsection{Cross sections of $^{87g}$Y}
\label{4.1.13}
The cumulative cross sections of $^{87g}$Y (T$_{1/2}$ = 79.8 h) contain the direct production, the contribution from the decay of $^{87m}$Y isomeric state (T$_{1/2}$ =13.37 h) and contributions from the decay of parents 87gZr (T$_{1/2}$ =1.6 h) and $^{87m}$Zr (T$_{1/2}$ =14 s, IT: 100 \%).  The same remark concerning the contributions of clusters and individual nucleons emissions as for $^{87m}$Y are valid. The agreement with the earlier experimental data and with TENDL-2013 is shown in Fig. 14.

\begin{figure}
\includegraphics[width=0.5\textwidth]{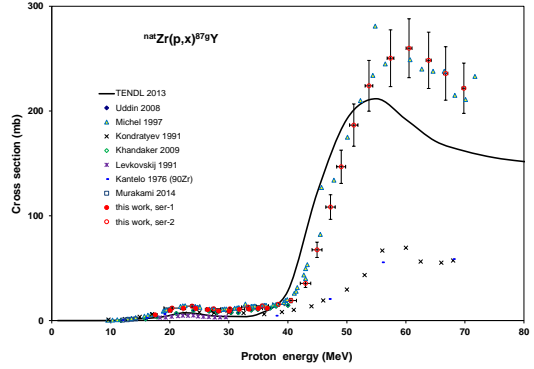}
\caption{Excitation function of the $^{nat}$Zr(p,x)$^{87g}$Y reaction}
\label{fig:14}       
\end{figure}

\subsubsection{Cross sections of $^{86}$Y}
\label{4.1.14}
The production cross sections for $^{86g}$Y (T$_{1/2}$ =14.74 h) contain the direct production and the contribution from the decay of short-lived isomeric state $^{86m}$Y (T$_{1/2}$ = 48 min, IT 99.3 \%). The contributions from the decay of $^{86}$Zr at high energies (see Fig. 11) were subtracted. The experimental and theoretical data are shown in Fig. 15.

\begin{figure}
\includegraphics[width=0.5\textwidth]{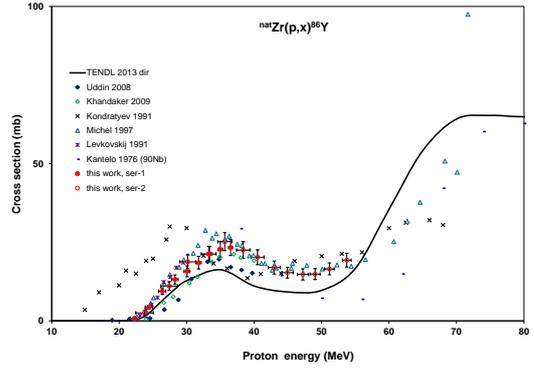}
\caption{Excitation function of the $^{nat}$Zr(p,x)$^{86}$Y reaction}
\label{fig:15}       
\end{figure}

\section{Integral yields}
\label{5}
Thick target yields (integrated yield for a given incident energy down to the reaction threshold) were calculated from fitted curves to our experimental cross section data. The results for physical yields (production rates) \citep{Bonardi} are presented in Figs. 16-18. Some earlier experimental thick target yield data found in the literature are also presented.

\begin{figure}
\includegraphics[width=0.5\textwidth]{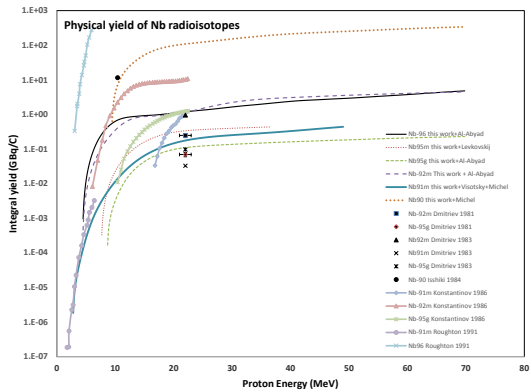}
\caption{Integral yields for production of $^{96}$Nb, $^{95}$Nb, $^{95g}$Nb, $^{92m}$Nb, $^{91m}$Nb, $^{90}$Nb deduced from the excitation functions}
\label{fig:16}       
\end{figure}

\begin{figure}
\includegraphics[width=0.5\textwidth]{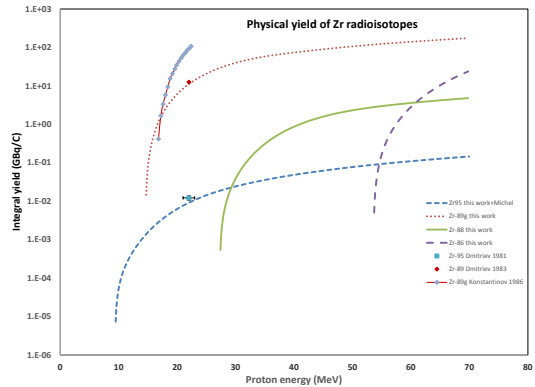}
\caption{Integral yields for production of  $^{95}$Zr,  $^{89}$Zr, $^{88}$Zr, $^{86}$Zr deduced from the excitation functions}
\label{fig:17}       
\end{figure}

\begin{figure}
\includegraphics[width=0.5\textwidth]{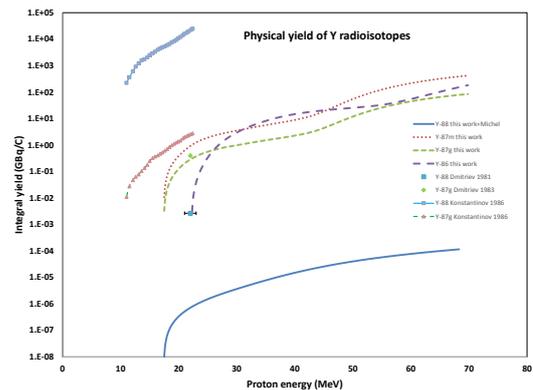}
\caption{Integral yields for production of  $^{88}$Y, $^{87m}$Y, $^{87g}$Y, $^{86}$Y  deduced from the excitation functions}
\label{fig:18}       
\end{figure}

\section{Production of medically relevant radioisotopes by charged particle induced reactions on Zr targets}
\label{6}
Among the studied activation products the radionuclides $^{90}$Nb, $^{95}$Nb, $^{89}$Zr, $^{88}$Y are of interest in nuclear medicine. The possible production routes for these nuclides using medium energy cyclotrons are shown in Table 6 and will be discussed for each nuclide separately with the aim to identify the possible role of proton induced reactions on Zr studied in this work. Nuclear reactions having low production yields, or resulting carrier added, low specific activity product are not included in the comparison.

\begin{table*}[t]
\tiny
\caption{Possible production routes of $^{90}$Nb, $^{95}$Nb, $^{89}$Zr and $^{88}$Y radioisotopes at medium energy cyclotrons (Enr. = enriched target material)}
\begin{center}
\begin{tabular}{|l|l|l|l|l|}
\hline
Product & Route & Reactions & Energy range & Target \\
\hline
\textbf{$^{90}$Nb} & direct & $^{90}$Zr(p,n)$^{90}$Nb & 8-20 & Enr. $^{90}$Zr \\
\hline
 & & $^{90}$Zr(d,2n)$^{90}$Nb & 12-28 & Enr. $^{90}$Zr \\
\hline
 & & $^{91}$Zr(p,2n)$^{90}$Nb & 20-9 & Enr. $^{91}$Zr \\
\hline
 & & $^{89}$Y($\alpha$,3n)$^{90}$Nb & 30-50 & $^{nat}$Y \\
\hline
 & & $^{89}$Y($^{3}$He,2n)$^{90}$Nb & 15-25 & $^{nat}$Y \\
\hline
\textbf{$^{95m}$Nb} & direct & $^{96}$Zr(p,2n)$^{95m}$Nb & 10-25 & Enr $^{96}$Zr \\
\hline
 & & $^{94}$Zr(d,n)$^{ 95m}$Nb & 5-15 & Enr $^{9}$Zr \\
\hline
 & & $^{96}$Zr(d,3n)$^{95m}$Nb & 12-30 & Enr $^{96}$Zr \\
\hline
\textbf{$^{89}$Zr} & direct & $^{89}$Y(p,n)$^{89}$Zr & 5-20 & $^{nat}$Y \\
\hline
 & & $^{89}$Y(d,2n)$^{89}$Zr & 10-30 & $^{nat}$Y \\
\hline
 & indirect & $^{90}$Zr(p,2n)$^{89}$Nb-$^{89}$Zr & 10-25 & Enr $^{90}$Zr \\
\hline
\textbf{$^{88}$Y} & direct & $^{nat}$Sr(p,xn)$^{88}$Y & 5-20 & $^{nat}$Sr \\
\hline
 & & $^{nat}$Sr(d,xn)$^{88}$Y & 5-20 & $^{nat}$Sr \\
\hline
 & & $^{nat}$Rb($\alpha$,xn)$^{88}$Y & 10-20 & $^{nat}$Rb \\
\hline
 & & $^{nat}$Zr(p,x)$^{88}$Y & 30-80 & $^{nat}$Zr \\
\hline
 & indirect & $^{nat}$Zr(p,x)$^{88}$Zr-$^{88}$Y & 30-80 & $^{nat}$Zr \\
\hline
 & & $^{89}$Y(p,x)$^{88}$Nb- $^{88}$Zr-$^{88}$Y & 15-35 & $^{nat}$Y \\
\hline
\end{tabular}
\end{center}
\end{table*}

\subsection{Production routes of $^{90}$Nb}
\label{6.1}
The $^{90}$Nb (T$_{1/2}$=23.35 h)  can be produced at low energy accelerators in various ways on Zr or Y targets ($^{90}$Zr(p,n)$^{90}$Nb,  $^{90}$Zr(d,2n)$^{90}$Nb,  $^{91}$Zr(p,2n)$^{90}$Nb,  $^{89}$Y($\alpha$,2n)$^{90}$Nb, $^{89}$Y($^3$He,n) $^{90}$Nb, see Table 6). 
Out of them only the $^{90}$Zr(p,n) and $^{89}$Y($^3$He,n)  reactions on enriched targets give products with high radionuclidic purity, as by limitations of the incident energy, reactions leading to Nb radionuclides with lower mass number can be suppressed (Q-value of $^{90}$Zr(p,2n) reaction is -17.002 MeV). In the case of natural Zr targets long-lived, higher mass Nb radionuclides are produced simultaneously e.g. $^{91m}$Nb (T$_{1/2}$=60.86 d) and $^{91g}$Nb (T$_{1/2}$ = 680 a) through the $^{91}$Zr(p,n) reaction. In Fig. 19 we reproduce the experimental cross section data of the  $^{90}$Zr(p,n) reaction (obtained on enriched $^{90}$Zr or derived from measurements on $^{nat}$Zr  (corrected by the TENDL  $^{91}$Zr(p,2n) data),  threshold of $^{91}$Zr(p,2n)$^{90}$Nb reaction is -14.244 MeV).  As it is shown the TENDL-2013 data reproduce well the experimental values for this reaction cross section.
In Fig. 20 for discussion we reproduce the excitation functions for $^{90}$Zr(d,2n)$^{90}$Nb and $^{89}$Y($\alpha$,2n)$^{90}$ Nb based on experimental data and for $^{90}$Zr(p,n)$^{90}$Nb and $^{89}$Y($^3$He,n) $^{90}$Nb reactions taken form the TENDL-2013 library. The deduced integral yields are shown in Fig. 21. It is clear from the cross sections and the integral yields, that the $^{90}$Zr(p,n)$^{90}$Nb  reaction is the method of choice  but requires highly enriched targets.

\begin{figure}
\includegraphics[width=0.5\textwidth]{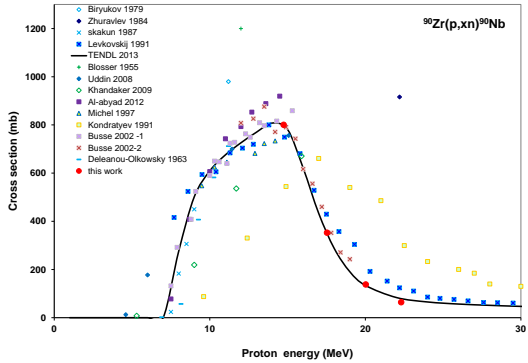}
\caption{Excitation function of the $^{90}$Zr(p,x)$^{90}$Nb reaction}
\label{fig:19}       
\end{figure}

\begin{figure}
\includegraphics[width=0.5\textwidth]{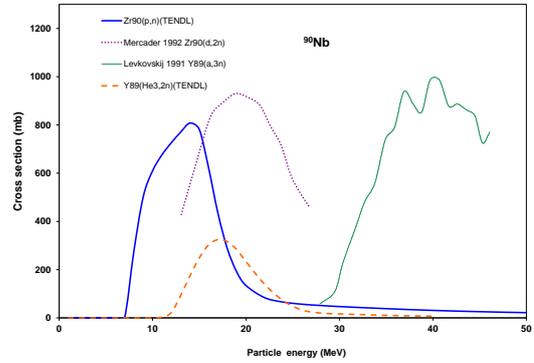}
\caption{Excitation function of the  $^{90}$Zr(p,n)$^{90}$Nb, $^{90}$Zr(d,2n)$^{90}$Nb, $^{89}$Y($\alpha$,3n) $^{90}$Nb and $^{89}$Y($^3$He,2n) $^{90}$Nb reactions}
\label{fig:20}       
\end{figure}

\begin{figure}
\includegraphics[width=0.5\textwidth]{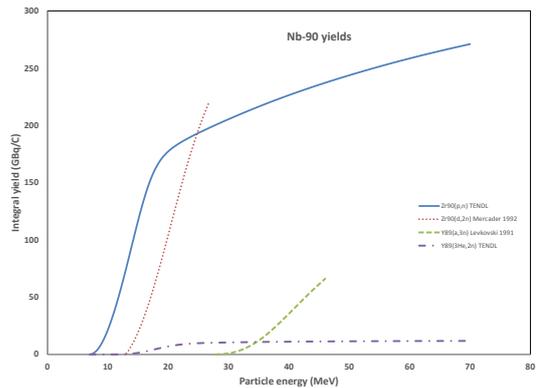}
\caption{Integral yields of the  $^{90}$Zr(p,n)$^{90}$Nb, $^{90}$Zr(d,2n)$^{90}$Nb, $^{89}$Y($\alpha$,3n) $^{90}$Nb and $^{89}$Y($^3$He,2n) $^{90}$Nb reactions}
\label{fig:21}       
\end{figure}

\subsection{Production routes of $^{95}$Nb}
\label{6.2}
The radionuclide $^{95}$Nb (T$_{1/2}$=3.61 d) can be produced with high specific activity and free from contaminants (if suitable cooling time is applied) with $^{96}$Zr(p,2n), $^{94}$Zr(d,n) and $^{96}$Zr(d,3n) reactions. The cross sections from experiments on enriched targets or the values derived from irradiations on $^{nat}$Zr targets are shown on Fig. 22 together with the data from the TENDL-2013 library. All of these reactions require highly enriched targets to obtain an end product with high specific activity and minimal contaminations. The $^{96}$Zr(p,2n) reaction  is the most  promising, considering both the yield and the required accelerator energy.

\begin{figure}
\includegraphics[width=0.5\textwidth]{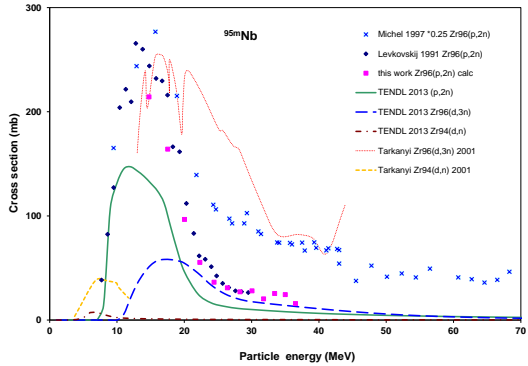}
\caption{Excitation function of  the $^{96}$Zr(p,2n), $^{94}$Zr(d,n) and $^{96}$Zr(d,3n) reactions}
\label{fig:22}       
\end{figure}

\subsection{Production routes of $^{89}$Zr}
\label{6.3}
The low and medium  energy production routes for  $^{89}$Zr (T$_{1/2}$=78.41 h) are $^{89}$Y(p,n)$^{89}$Zr and $^{89}$Y(d,2n)$^{89}$Zr as direct routes and $^{90}$Zr(p,2n)$^{89}$Nb-$^{89}$Zr as indirect route. A large number of experimental data sets exist for the direct (p,n) and (d,2n) reactions (see Fig. 23 and 24, the experimental data were taken from EXFOR).  Both reactions result in non-carrier added (nca) product and the yttrium has only one stable isotope. The (p,n) route hence seems to be the most  beneficial, both in yield and the required accelerator characteristics. In principle $^{89}$Zr can be produced carrier free indirectly by using Zr targets through the $^{90}$Zr(p,2n)$^{89}$Nb-$^{89}$Zr reaction. No experimental data are available for this reaction. The theoretical data from TENDL-2013 are shown in Fig. 23.
When considering integral yields the $^{90}$Zr(p,2n)$^{89}$Nb-$^{89}$Zr  indirect production route  is the most productive, but it requires highly enriched targets and only short irradiations are possible due to the short half-life of the isomeric states of the $^{89}$Nb ($^{89m}$Nb T$_{1/2}$ = 66 min, $^{89g}$Nb T$_{1/2}$= 2.13 h) and the need to separate the Nb from the Zr target to obtain high specific activity NCA end-product.

\begin{figure}
\includegraphics[width=0.5\textwidth]{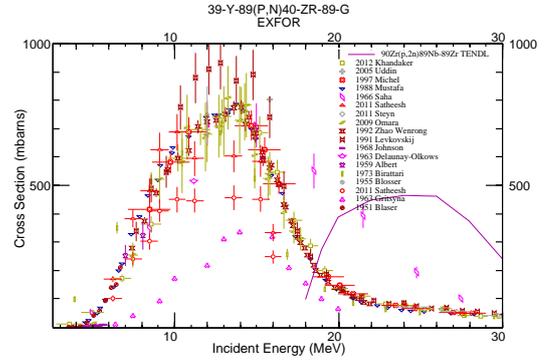}
\caption{Excitation function of  the $^{89}$Y(p,n) $^{89}$Zr reaction, plotted together with the $^{90}$Zr(p,2n)$^{89}$Nb	reaction cross section also producing the $^{89}$Zr isotope}
\label{fig:23}       
\end{figure}

\begin{figure}
\includegraphics[width=0.5\textwidth]{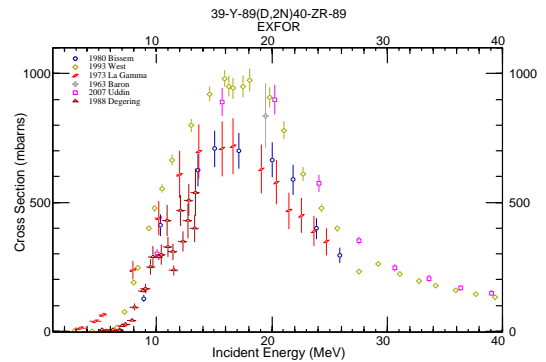}
\caption{Excitation function of the $^{89}$Y(d,2n) $^{89}$Zr reaction}
\label{fig:24}       
\end{figure}

\subsection{Production routes of $^{88}$Y}
\label{6.4}
The direct production of $^{88}$Y (T$_{1/2}$ = 106.627 d) is possible through the $^{nat}$Sr(p,xn), $^{nat}$Sr(d,xn),  $^{nat}$Rb($\alpha$,xn),  $^{nat}$Zr(p,x) and $^{90}$Zr(p,x) reactions. The excitation functions for the direct production are shown in Figs. 25-26 (the data were taken from EXFOR or from TENDL-2013 for the (d,xn) reaction) and in Fig. 12. By using $^{nat}$Zr(p,x) the cross section is low and long-lived $^{91}$Y (58.51 d) is produced simultaneously. To get high radionuclidic purity highly enriched $^{90}$Zr should be used. 
As Rb has only two stable isotopes ($^{85}$Rb:72.165 \% and $^{87}$Rb:27.83\%) and $^{90}$Y with a significantly shorter half-life is produced by ($\alpha$,n) reaction, natural Rb targets can be used. However the limitation on incoming $\gamma$-energy is important for avoiding production of stable $^{89}$Y through $^{87}$Rb($\alpha$,2n), which would decrease specific activity. The maximum production  cross section  on  natRb will hence be only around 230 mb.
The indirect production routes include the $^{nat}$Zr(p,x)$^{88}$Zr-$^{88}$Y, $^{90}$Zr(p,x)$^{88}$Zr-$^{88}$Y,  $^{89}$Y(p,2n) $^{88}$Zr-$^{88}$Y reactions. The excitation functions are shown in Fig. 10 and Fig. 27 (the experimental data for $^{89}$Y were taken from EXFOR).
By using Zr target the production requires highly enriched $^{90}$Zr material. In case of natural composition other long-lived Zr radioisotopes as $^{93Z}$r (T$_{1/2}$= 1.61106 a) and $^{95}$Zr T$_{1/2}$= 64.032 d) are produced simultaneously, followed by Y decay product. In case $^{90}$Zr target the $^{88}$Y is produced directly, simultaneously with $^{88}$Zr via (p,2pxn) reaction, which can be separated after EOB (the other Y radio-products are short-lived).
Comparison of the two production routes shows, that the (p,2n) reaction on monoisotopic yttrium has advantages as no enriched targets ate required, and the required beam energy is lower up to 30 MeV (in the range of commercial medium energy cyclotrons).

\begin{figure}
\includegraphics[width=0.5\textwidth]{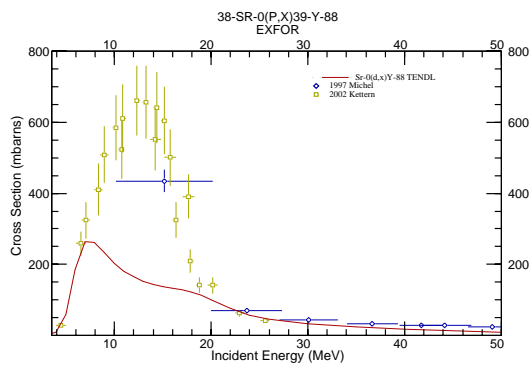}
\caption{Excitation function of  the natSr(p,x)$^{88}$Y and  natSr(d,x)$^{88}$Y reaction}
\label{fig:25}       
\end{figure}

\begin{figure}
\includegraphics[width=0.5\textwidth]{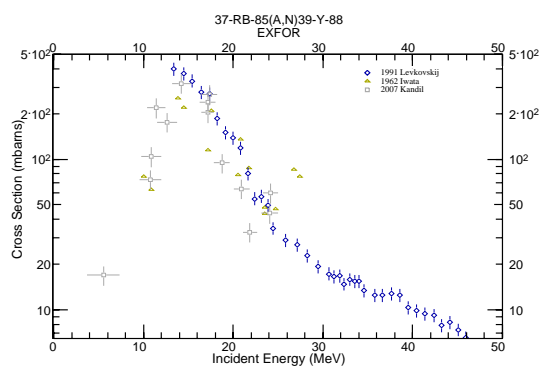}
\caption{Excitation function of $^{85}$Rb(,n) reaction}
\label{fig:26}       
\end{figure}

\begin{figure}
\includegraphics[width=0.5\textwidth]{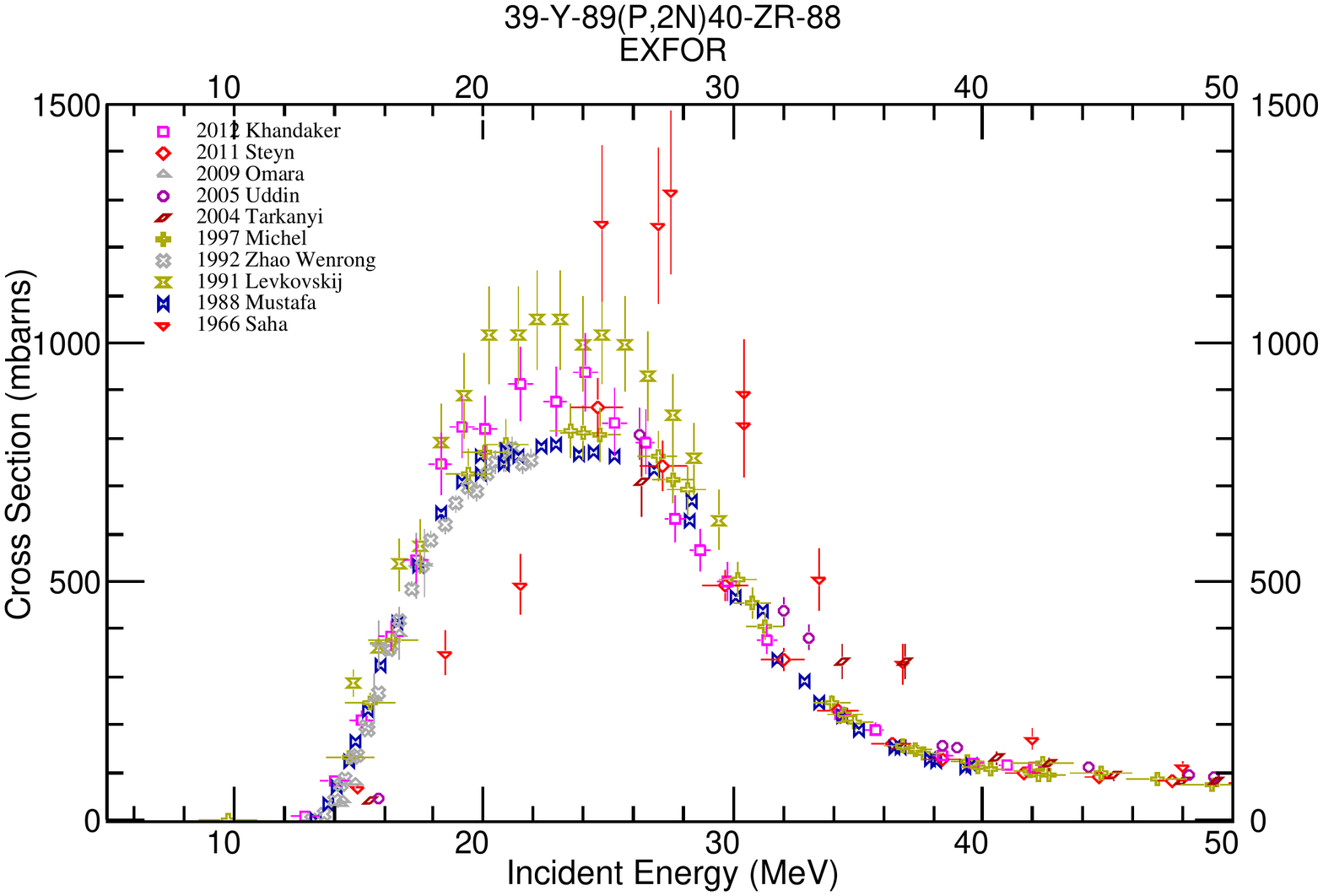}
\caption{27.	Excitation function of the $^{89}$Yp,2n)$^{88}$Zr reaction}
\label{fig:27}       
\end{figure}

\section{Summary}
\label{7.}
We present experimental activation cross sections for production of $^{96}$Nb, $^{95}$Nb, $^{95g}$Nb, $^{92m}$Nb, $^{91m}$Nb, $^{90}$Nb, $^{95}$Zr, $^{89}$Zr, $^{88}$Zr, $^{86}$Zr, $^{88}$Y, $^{87m}$Y, $^{87g}$Y, $^{86}$Y  on zirconium measured up to 70 MeV proton energy. The agreement between the new and the earlier experimental data (except for a few cases) is acceptable. TALYS 1.4 based model results in TENDL-2013 library describe well the experimental results. For production of the medically relevant radionuclides $^{90}$Nb, $^{95}$Nb, $^{89}$Zr, $^{88}$Y   different alternative routes were compared and discussed. For production of $^{90}$Nb and $^{95}$Nb the proton induced reactions on enriched $^{90}$Zr and $^{96}$Zr, respectively, are preferred compared to other routes. For production of $^{89}$Zr and $^{88}$Y the proton and deuteron induced reactions on $^{nat}$Y from all points of view are more practical.

\section{Acknowledgements}
\label{}
This work was partly performed in the frame of the HAS-FWO Vlaanderen (Hungary-Belgium) project as well as in cooperation with the Tohoku University, Sendai, Japan. The authors acknowledge the support of the research project and the respective institutions in providing the beam time and experimental facilities.
 



\bibliographystyle{elsarticle-harv}
\bibliography{Zrp}







\end{document}